\newif\ifsubmode
\submodefalse

\newif\ifprintfig
\printfigtrue

\ifsubmode
  \documentstyle[12pt,aasms4,epsf]{article}
  \received{}
  \accepted{}
  \journalid{}{}
  \articleid{}{}
\else
  \documentstyle[11pt,aaspp4,epsf]{article}
  \slugcomment{{\it The Astronomical Journal, submitted 10/27/1999}}
\fi

\lefthead{van der Marel et al.}
\righthead{The velocity and mass distribution of clusters}

\newcommand{\etal}{{et al.~}}

\newcommand{\lta}{\lesssim}
\newcommand{\gta}{\gtrsim}

\newcommand{\kms}{\>{\rm km}\,{\rm s}^{-1}}

\newcommand{\Mpc}{\>{\rm Mpc}}
\newcommand{\Msun}{\>{\rm M_{\odot}}}

\newcommand{\Rt}{{\tilde R}}

\newcommand{\vt}{{\tilde v}}

\newcommand{\fint}{f_{\rm i}}

\begin{document}

\title{The velocity and mass distribution of clusters of galaxies\\
from the CNOC1 cluster redshift survey}

\author{Roeland P.~van der Marel}
\affil{Space Telescope Science Institute, 3700 San Martin Drive, 
Baltimore, MD 21218}

\author{John Magorrian}
\affil{Institute of Astronomy, Madingley Road, Cambridge CB3 0HA, 
United Kingdom}

\author{Ray G.~Carlberg, H.~K.~C.~Yee}
\affil{Department of Astronomy, University of Toronto, Toronto ON, M5S 1A7,
Canada}

\author{E.~Ellingson}
\affil{Center for Astrophysics \& Space Astronomy, University of Colorado, 
Boulder, CO 80309-0389}



\ifsubmode\else
\clearpage\fi


\ifsubmode\else
\baselineskip=14pt
\fi


\begin{abstract}
In the context of the CNOC1 cluster survey, redshifts were obtained
for galaxies in 16 clusters. The resulting sample is ideally suited
for an analysis of the internal velocity and mass distribution of
clusters. Previous analyses of this dataset used the Jeans equation to
model the projected velocity dispersion profile. However, the results
of such an analysis always yield a strong degeneracy between the mass
density profile and the velocity dispersion anisotropy profile. Here
we analyze the full $(R,v)$ dataset of galaxy positions and velocities
in an attempt to break this degeneracy.

We build an `ensemble cluster' from the individual clusters under the
assumption that they form a homologous sequence; if clusters are not
homologous then our results are probably still valid in an average
sense. To interpret the data we study a one-parameter family of
spherical models with different constant velocity dispersion
anisotropy, chosen to all provide the same acceptable fit to the
projected velocity dispersion profile. The best-fit model is sought
using a variety of statistics, including the likelihood of the
dataset, and the shape and Gauss-Hermite moments of the grand-total
velocity histogram. The confidence regions and goodness-of-fit for the
best-fit model are determined using Monte-Carlo simulations. Although
the results of our analysis depend slightly on which statistic is used
to judge the models, all statistics agree that the best-fit model is
close to isotropic. For none of the statistics does the 1-$\sigma$
confidence region extend below $\sigma_r / \sigma_t = 0.74$, or above
$\sigma_r / \sigma_t = 1.05$.  This result derives primarily from the
fact that the observed grand-total velocity histogram is close to
Gaussian, which is not expected to be the case for a strongly
anisotropic model.

The best-fitting models have a mass-to-number-density ratio that is
approximately independent of radius over the range constrained by the
data. They also have a mass-density profile that is consistent with
the dark matter halo profile advocated by Navarro, Frenk \& White, in
terms of both the profile shape and the characteristic scale
length. This adds important new weight to the evidence that clusters
do indeed follow this proposed universal mass density profile.

We present a detailed discussion of a number of possible uncertainties
in our analysis, including our treatment of interlopers and brightest
cluster galaxies, our use of a restricted one-parameter family of
distribution functions, our use of spherical models for what is in
reality an ensemble of non-spherical clusters, and our assumption that
clusters form a homologous set. These issues all constitute important
approximations in our analysis. However, none of the tests that we
have done indicates that these approximations influence our results at
a significant level.
\end{abstract}



\clearpage


\section{Introduction}
\label{s:intro}

Determinations of the internal velocity and mass distribution of
galaxy clusters are of great value, since they have the potential to
constrain both the main cosmological parameters and scenarios for
large-scale structure formation (e.g., Crone, Evrard \& Richstone
1994; Cole \& Lacey 1996; de Theije, van Kampen \& Slijkhuis 1998,
1999). A recent development on this problem has been the prediction
from cosmological simulations that dark matter halos have a universal
density profile (Navarro, Frenk \& White 1997; hereafter NFW). Testing
the validity of this prediction is an important goal for any study of
cluster structure. The traditional way to study these issues is to
construct dynamical equilibrium models for the redshift measurements
of individual cluster galaxies. Alternative approaches to infer the
mass distribution of clusters are to use X-ray observations (e.g.,
Allen 1998; Hughes 1998), measurements of the weak- or strong-lensing
of background sources (e.g., Bartelmann \& Narayan 1995), or caustics
in redshift space (Geller, Diaferio \& Kurtz 1999; Diaferio
1999). Each of the different methods has its own uncertainties and
biases, and results from different methods are often found to
differ. The causes and magnitudes of these differences remain a hot
topic of debate (e.g., Lewis \etal 1999 and references therein).

Here we focus attention on the dynamical analysis of galaxy
redshifts.\footnote{Dynamical models of clusters of galaxies based on
the Boltzmann equation (or its integrals, the Jeans equations and the
virial theorem) always assume implicitly that the system is in
equilibrium. While this is not strictly true for a cluster of
galaxies, numerical simulations show that cluster evolution proceeds
through a series of quasi-equilibrium states that satisfy the
Boltzmann equation in an approximate sense (Natarajan, Hjorth, \& van
Kampen 1997).}  Analyses of this type have nearly always been
restricted to modeling of the projected velocity dispersion profile
$\sigma(R)$ (e.g., Kent \& Gunn 1982; Merritt 1987), often using the
Jeans equation (e.g., Solanes \& Salavdor-Sole 1990; den Hartog \&
Katgert 1996; Natarajan \& Kneib 1996). The main limitation of such
analyses is that there are two unknown functions of one variable, the
mass density profile $\rho(r)$ and the velocity dispersion anisotropy
profile $\beta(r)$, that are both constrained by only one function of
one variable, $\sigma(R)$. As a result, there is always a degenerate
set of models with different $\rho(r)$ and $\beta(r)$ that can all fit
the observations equally well (Binney \& Mamon 1982). Attempts are
often made to break this degeneracy by assuming that either one of
$\rho(r)$ or $\beta(r)$ is known, so that the other can be determined.
Popular assumptions are that the velocity distribution is isotropic,
or that the mass distribution can be calculated from the observed
number density by the assumption of a constant mass-to-number density
(or mass-to-light) ratio. However, neither of these assumptions has a
strong physical justification, so these approaches do not remove the
underlying degeneracy.

As a result of steady improvements in instrumentation, and in
particular the advent of efficient multi-object spectrographs, the
available redshift samples for galaxy clusters have been steadily
increasing in size. Important recent datasets are the ENACS (ESO
Nearby Abell Cluster Survey) (e.g., Katgert \etal 1996) and the CNOC1
(Canadian Network for Observational Cosmology) cluster redshift
survey\footnote{We refer to the CNOC cluster redshift survey as
`CNOC1', to distinguish it from the subsequent CNOC2 field galaxy
redshift survey (e.g., Lin \etal 1999).} (e.g., Yee, Ellingson \&
Carlberg 1996). With improved statistics it should become possible to
extract not only velocity dispersion profiles from the data, but also
higher order moments that describe deviations of the observed velocity
histograms from a Gaussian shape (Zabludoff, Franx \& Geller
1993). Such measurements have the potential to break the degeneracy
between the mass and velocity distribution, as emphasized in this
context by, e.g., Merritt (1987; 1993) and Merritt \& Saha (1993).

Here we focus on the data from the CNOC1 cluster survey. Redshifts
were obtained for galaxies in 16 clusters at $z = 0.17$--$0.55$,
selected on the basis of their X-ray luminosity.  Previous analyses of
these data were all based either on use of the virial theorem
(Carlberg \etal 1996; hereafter C96) or the Jeans equation (Carlberg
\etal 1997a,b,c; hereafter C97a,b,c). These analyses yielded the
important result that if cluster of galaxies have a velocity
distribution that is not too far from isotropic, then the CNOC1 data
are consistent with clusters having an approximately constant
mass-to-number-density ratio, and a mass-density profile that is
approximately of the NFW form with a scale radius that is consistent
with predictions from cosmological simulations. However, because of
the degeneracy intrinsic to the Jeans equation, models with
significant anisotropies and very different mass profiles can fit the
projected velocity dispersion profile equally well. So the important
question is: can the range of allowed models for the CNOC1 data be
reduced by considering not only the projected velocity dispersion
profile, but also the entire $(R,v)$ dataset and the implied
higher-order velocity moments? This question is the main topic of the
present paper. To answer it, we present a new analysis of the CNOC1
dataset, in part similar in spirit to that suggested by Merritt \&
Saha (1993), but with some differences (see \S\ref{ss:discDF} below).

In \S\ref{s:jeansmodels} we first repeat some of the Jeans modeling of
$\sigma(R)$, but using a different approach from that employed by
C97a,b,c. The results not only provide an illustration of the
degeneracies involved in the modeling, but also serve as a useful
starting point for the more detailed analysis. In \S\ref{s:DFmodels}
we analyze the full $(R,v)$ dataset. We seek a best-fit model using a
variety of statistics, including the likelihood of the dataset and the
shape and Gauss-Hermite moments of the grand-total velocity
histogram. The confidence regions and goodness-of-fit for the best-fit
model are estimated using Monte-Carlo simulations. In
\S\ref{ss:uncertainties} we discuss possible uncertainties
in our analysis, and how robust the conclusions of our analysis are in
view of these uncertainties. In \S\ref{s:discconc} we present and
discuss the final conclusions. In Appendix A we compare our modeling
approach to that of Ramirez \& De Souza (1998) and Ramirez, De Souza
\& Schade (1999), who recently used a more approximate method to
constrain the orbital anisotropy of cluster galaxies from observed
cluster velocity histograms (including that for the CNOC1 sample).

\section{Jeans equation models}
\label{s:jeansmodels}

In the CNOC1 cluster survey redshifts were obtained for galaxies in 16
different clusters. The characteristic size and velocity of each
cluster can be quantified by, e.g., the radius $r_{200}$ inside which
the average mass density equals $200$ times the critical density of
the Universe (at the given redshift), and the line-of-sight velocity
dispersion $\sigma$. For our dynamical analysis we treat the data
similarly as in C97b: we assume that clusters form a homologous
sequence, each with identical structure in dimensionless units. To
study this dimensionless structure we consider the dataset that
contains for each galaxy in the survey the quantities $\Rt \equiv
R/r_{200}$ and $\vt \equiv v/\sigma$, where $R$ is the projected
distance from the cluster center, $v$ is the observed line-of-sight
velocity with respect to the systemic cluster velocity, and $r_{200}$
and $\sigma$ are taken from the analysis of C97b. One may consider the
$(\Rt,\vt)$ as quantities drawn from `the ensemble cluster'. The
advantage of combining data from different clusters is that it reduces
the influence of, e.g., substructure and non-sphericity on the
properties of the dataset. With this approach, spherical equilibrium
models are likely to provide an adequate description of the data (see
\S\ref{ss:axisym} and \S\ref{ss:homology} below).

Hereafter we write $(R,v)$ instead of $(\Rt,\vt)$, with the
understanding that all quantities discussed are dimensionless unless
otherwise noted. In our dynamical analysis we also work in
dimensionless units. To this end we adopt a unit of mass
\begin{equation}
  M_{\rm u} = (2.3252 \times 10^{14} \Msun) \>
              (r_{200}/\Mpc) \>
              (\sigma/[10^3 \kms])^2
\label{Munit}
\end{equation}
chosen to set the gravitational constant to $G=1$.

The projected galaxy number density profile $\Sigma(R)$ of the CNOC1
ensemble cluster was derived in C97b, with corrections for both the
non-uniform sampling of the clusters (see Yee \etal 1996) and
interloper contamination. Figure~\ref{f:projnumden} shows the result;
$\Sigma$ is the number of galaxies per unit area with K-corrected
absolute Gunn $r$-band magnitude $M_r < -18.5$,\footnote{The value of
the absolute magnitude limit assumes a Hubble constant $H_0 = 100 \kms
\Mpc^{-1}$. This is the only quantity in this paper that depends on
$H_0$, because our entire analysis is performed in dimensionless
units.} normalized to unity over the circular region with radius
$r_{200}$. For the dynamical modeling we have fitted to the data a
smooth function of the form
\begin{equation}
  \Sigma(R) = \Sigma_b \> 2^{{\beta-\gamma}\over\alpha} \> 
         (R/b)^{-\gamma} \>
         [1 + (R/b)^\alpha]^{-{{\beta-\gamma}\over\alpha}} \>
         [1 + (R/c)^\delta]^{-{{\epsilon-\beta}\over\delta}} .
\label{projnumden}
\end{equation}
This somewhat arbitrary parameterization is similar to the `nuker-law'
advocated by, e.g., Byun \etal (1996), but is more general in having
two power-law breaks rather than one. A good fit was obtained with
$\Sigma_b = 1.03$, $b = 0.21$, $c = 1.32$, $\alpha = 2.97$, $\beta =
1.51$, $\gamma = 0.65$, $\delta = 4.00$, and $\epsilon = 3.00$. The
parameterized fit is shown in Figure~\ref{f:projnumden}. The fit to
the data is statistically acceptable, but note that the fit is not
unique; the parameters of equation~(\ref{projnumden}) are strongly
correlated, and are not all equally well constrained. C97c showed that
the projection of a profile such as that advocated by NFW also
provides an acceptable fit to the data.

To interpret the dynamics of the ensemble cluster we first model the
projected velocity dispersion profile using the Jeans equation for a
spherical system. The software used for this was similar to that
discussed in van der Marel (1994). The intrinsic galaxy number density
profile $\nu(r)$ follows from the projected galaxy number density
profile $\Sigma(R)$ by solution of the relevant Abel transform
equation. An ansatz is made for the mass density profile $\rho(r)$ of
the cluster, and from this the gravitational potential and
gravitational force are calculated. The Jeans equation is then solved
for the intrinsic velocity dispersions of the system (the tangential
velocity dispersion $\sigma_t = \sigma_{\theta} = \sigma_{\phi}$ and
the radial velocity dispersion $\sigma_r$), for some velocity
dispersion anisotropy profile $\beta(r)$, where $\beta \equiv 1 -
\sigma^2_t/\sigma^2_r$. The intrinsic velocity dispersion components
in the direction to the observer are then weighted with the intrinsic
galaxy number density and projected along the line of sight to yield
the projected line-of-sight velocity dispersion profile $\sigma(R)$,
which can be compared to observations.

As discussed in \S\ref{s:intro}, there is always a degenerate set of
models with different $\rho(r)$ and $\beta(r)$ that all fit
$\sigma(R)$ equally well. The goal of the present paper is to see to
what extent this degeneracy can be broken by modeling not only the
projected velocity dispersion profile, but also the individual $(R,v)$
data points (and thus indirectly the higher order velocity
moments). We do not wish to be too ambitious and therefore restrict
ourselves to a one-parameter family of models, namely those in which
the velocity dispersion anisotropy is independent of radius, $\beta(r)
= {\rm constant}$. There is a unique mass density profile $\rho(r)$
for each constant $\beta$, such that models with different $\beta$ all
provide an identical fit to the observations. Instead of attempting to
infer non-parametrically the optimum $\rho(r)$ for each $\beta$, we
adopt a simple three-parameter family of models among which we seek
the one that fits best. We set
\begin{equation}
  \rho(r) = \rho_0 \> (r/a)^{-\xi} \> [1+(r/a)]^{\xi-3} .
\label{massdens}
\end{equation}
The parameters $\rho_0$ and $a$ set the scale of the mass distribution
in density and length, while $\xi$ is the logarithmic power-law slope
near the center. For $\xi=1$ this mass density is of the form
advocated by NFW; for $\xi=0$ the mass density has a homogeneous core.

The projected velocity dispersion profile $\sigma(R)$ of the CNOC1
ensemble cluster was derived in C97b, with appropriate correction for
interloper contamination. The dispersion profile that we have used in
our analysis was obtained with a similar but slightly updated
treatment of the data, and is shown in Figure~\ref{f:dispprof}. The
inferred profile is mildly different from that analyzed in C97b, but
is equivalent in a statistical sense. Eleven constant-$\beta$ models
were constructed to interpret the data, with $\beta$ chosen such that
the ratio $\sigma_r / \sigma_t$ was sampled logarithmically between
$1/3$ and $3$. For each $\beta$ we determined the $\rho(r)$ for which
the model predictions best fit the data. To this end we used a grid in
the $(a,\xi)$ parameter space. For each $(a,\xi)$ we calculated the
shape of the predicted velocity dispersion profile, and the density
normalization $\rho_0$ that yields the best fit to the observed
velocity dispersion profile in a $\chi^2$ sense. We then sought the
minimum $\chi^2$ over the $(a,\xi)$ parameter space to find the
best-fitting mass density profile $\rho(r)$ for the given $\beta$.
The velocity dispersion profiles predicted by these best-fit models
are shown as curves in Figure~\ref{f:dispprof}. Apart from their small
radii behavior (where there are no strong constraints from the data)
the model predictions are similar for the different values of $\beta$,
as expected based on the degeneracy of the problem. The small
differences between the various models are due to the fact that a
parameterized form was used for the mass density $\rho(r)$; the true
$\rho(r)$ that best fits the data may not be exactly of the adopted
form. However, the fits to the data in Figure~\ref{f:dispprof} are all
statistically acceptable, as judged by the $\chi^2$ of the fit. So
even though a non-parameterized modeling approach would likely yield
mass densities $\rho(r)$ that differ from the parameterized form
adopted here, the differences would not be statistically
significant. This provides a posteriori justification for the use of
the adopted mass density parameterization.

Figure~\ref{f:massparam} shows the parameters $\rho_0$, $a$ and $\xi$
of the best-fitting mass density $\rho(r)$ as a function of the
anisotropy $\sigma_r / \sigma_t$. Solid dots indicate the models for
which the predictions are shown in Figure~\ref{f:dispprof}. The
profiles $\rho(r)$ for these models are shown in the top right panel
of Figure~\ref{f:massprof}. As the value of $\sigma_r /
\sigma_t$ is increased, the scale radius $a$ increases while the
small radii slope $\xi$ decreases. Consequently, tangentially
anisotropic models have higher mass densities and steeper mass density
profiles at small radii than radially anisotropic models. Very
radially anisotropic models formally attain their best fit for $\xi <
0$, i.e., with decreasing mass densities near the center. This seems
implausible for various reasons, and we therefore fixed $\xi = 0$ for
these models. This has little effect on the quality of the fit to the
observed velocity dispersion profile, cf.~Figure~\ref{f:dispprof}.

The bottom right panel of Figure~\ref{f:massprof} shows the enclosed
mass profile $M(r)$ for the models. The profile shapes depend strongly
on the anisotropy, with tangentially anisotropic models being more
centrally concentrated than radially anisotropic
models. Interestingly, $M_{200}$, the enclosed mass within $r_{200}$
($\log(r)=0$ in dimensionless units), is virtually independent of the
assumed anisotropy.\footnote{Efstathiou, Ellis, \& Carter (1980)
studied the case of test-particles with number density $\nu(r) \propto
r^{-3}$ in a spherical isothermal potential. The enclosed mass $M(r)$
is then strictly independent of $\beta$ for all radii. The present
result is different in that $M(r)$ for the models studied here
does depend strongly on $\beta$ for $r \ll r_{200}$ and $r \gg
r_{200}$, but not for $r \approx r_{200}$.} This was noticed
previously by C97b, and significantly reduces the uncertainty in
estimates of the cosmological mass density $\Omega$ from cluster
mass-to-light ratios using Oort's method.

The top left panel of Figure~\ref{f:massprof} shows the number density
profile $\nu(r)$ of the models, as obtained by deprojection of the
projected profile shown in Figure~\ref{f:projnumden}. When combined
with the mass density profile $\rho(r)$ for each model one obtains the
mass-to-number-density ratio $\rho/\nu(r)$, which is shown in the
bottom left panel of Figure~\ref{f:massprof}. This ratio decreases
with radius for tangentially anisotropic models, and increases with
radius for radially anisotropic models. The isotropic model (heavy
curve) has a mass-to-number-density ratio $\rho/\nu(r)$ that is very
nearly independent of radius. Note that the logarithmic slopes of the
adopted $\nu(r)$ and $\rho(r)$ differ at asymptotically small and
large radii (for $r \rightarrow 0$: $\nu(r) \propto r^{-1.65}$ and
$\rho(r) \propto r^{-\xi}$; for $r \rightarrow \infty$: $\nu(r)
\propto r^{-4}$ and $\rho(r) \propto r^{-3}$). So it is not an
intrinsic property of the adopted parameterizations that isotropic
models have approximately constant $\rho/\nu(r)$ over the region where
this quantity is constrained by the data.

If the galaxy luminosity function of clusters is independent of
position, then the mass-to-light ratio $M/L$ is directly proportional
to the mass-to-number-density ratio $\rho/\nu(r)$. In physical units,
$M/L$ is given by the dimensionless function $\rho/\nu(r)$ shown in
Figure~\ref{f:massprof}, multiplied by $M_{\rm u}/L_{200}$. Here
$M_{\rm u}$ is as given in equation~(\ref{Munit}), and $L_{200}$ is
the total luminosity inside a projected radius of $r_{200}$. The
quantities $r_{200}$, $\sigma$ and $L_{200}$ for the individual
clusters of the CNOC1 redshift survey are listed in Table~4 of C96 and
Table~1 of C97b.

NFW have argued that cosmological simulations yield a more-or-less
universal mass density profile $\rho(r)$ for dark matter halos, given
by equation~(\ref{massdens}) with $\xi=1$. Our models generate such a
profile if $\log(\sigma_r / \sigma_t) = -0.06$, i.e., $\sigma_r /
\sigma_t \approx 0.87$ (cf.~Figure~\ref{f:massparam}). This model is 
close to isotropic, but has mild tangential anisotropy. Like the
isotropic model, it has an approximately constant
mass-to-number-density ratio (cf.~Figure~\ref{f:massprof}). The
logarithmic slope of $\rho(r)$ at the last data point ($R = 1.5$)
equals $2.7$. So even though this model is consistent with an NFW
profile, the data do not actually allow us to test whether the mass
density slope converges to $3$ at large radii, as in the NFW profile.
Figure~\ref{f:massparam} shows that the model with $\log(\sigma_r /
\sigma_t) = -0.06$ has scale length $a = 0.24$ (as before, in units of
$r_{200}$). This is in excellent agreement with the predictions of
NFW's cosmological simulations, which predict $a = 0.20$ for an
$\Omega = 0.2$ open cold dark matter model and $a = 0.26$ for a flat
$\Omega = 0.2$ model (cf.~C97c). The Jeans modeling therefore shows
that the CNOC1 data are consistent with both an NFW mass density
profile and a constant mass-to-number-density ratio for clusters of
galaxies, but only if the unknown velocity dispersion anisotropy has a
particular value that is close to isotropic. Figures~\ref{f:dispprof}
and~\ref{f:massprof} show that models with other properties can fit
the data equally well. To break this degeneracy we proceed with a more
sophisticated analysis that uses the entire $(R,v)$ dataset, instead
of just the projected velocity dispersion profile.
 
\section{Distribution function models}
\label{s:DFmodels}

\subsection{Model calculation}
\label{ss:DFcalc}

To interpret the velocities of individual galaxies in the CNOC1
ensemble cluster we need to construct models based on phase space
distribution functions (DFs)\footnote{Ramirez \& De Souza (1998) and
Ramirez, De Souza \& Schade (1999) bypassed the calculation of DFs in
their modeling of the projected velocity distributions of clusters by
making a number of simplifying assumptions, and they used the
resulting kinematic models to draw conclusions about the velocity
dispersion anisotropy of cluster galaxies of different morphological
types. Appendix~A presents a quantitative assesment of the validity of
their approach and conclusions.}. As before, we restrict ourselves to
models with constant anisotropy $\beta$. For each $\beta$, we fix the
three-dimensional galaxy number density $\nu(r)$ and mass density
$\rho(r)$ to those calculated in \S\ref{s:jeansmodels} from the Jeans
models. Fixing the anisotropy $\beta$ does not uniquely determine the
DF. For given $\nu(r)$ and $\rho(r)$ there are infinitely many DFs
that all predict the same second order velocity moments; a DF is
determined uniquely only after specification of all its velocity
moments (e.g., Dejonghe 1986). Here we do not seek to derive the full
set of DFs that can generate a model with a given anisotropy, but
instead we are satisfied to find just one. To achieve this we make a
simple ansatz for the DF: $f = f_{\beta}(E,L) \equiv g_{\beta}(E)
L^{-2 \beta}$, where $E$ and $L$ are the binding energy and angular
momentum per unit mass. Such models have fixed constant anisotropy
$\beta$ (e.g., H\'enon 1973; Kent \& Gunn 1982; Cuddeford 1991). The
problem lies in finding the function $g_{\beta}(E)$ that generates the
required $\nu(r)$ in the given $\rho(r)$. The number-density profile
is given by
\begin{equation}
  \nu(r) \> \equiv \> 
            \int f(E,L) \> {\rm d}^3 {\b v}
         \> = \> 
            2^{3/2-\beta} \> \pi^{3/2} \> r^{-2\beta} \> 
            {{\Gamma(1-\beta)}\over{\Gamma({3\over2}-\beta)}} \>
            \int_0^{\psi(r)} g_\beta(E)
            [\psi(r)-E]^{1/2-\beta} \> {\rm d}E .
\label{FIE}
\end{equation}
Here $\psi(r) = E - {1\over2}v^2$ is the relative gravitational
potential, which is uniquely determined by the mass density
distribution $\rho(r)$ through Poisson's equation (e.g., Binney \&
Tremaine 1987).

For each fixed value of $\beta$ we solve equation~(\ref{FIE}) for
$g_\beta(E)$ using the penalized likelihood method described in
Appendix~C of Magorrian \& Tremaine (1999). This generally yields a
solution for which the predicted $\nu(r)$ agrees to within $\lta 1$\%
with the $\nu(r)$ inferred from the data. Only for radially
anisotropic models with $\sigma_r / \sigma_t \gta 2$ could we not
achieve such good agreement. This is not due to a numerical flaw in
our approach, but indicates that such radially anisotropic DFs cannot
produce central density cusps that are as shallow as observed
(Figure~\ref{f:projnumden}). This by itself can be taken as evidence
against these models. However, the models fail only at radii $r \lta
0.1$, where the constraints from the data are not very strong (due to
small-number statistics). Because of this, we have not excluded
radially anisotropic models with $\sigma_r / \sigma_t \gta 2$ from our
analysis. Instead we use for them the best, albeit imperfect, solution
to equation~(\ref{FIE}). This has little impact on our final
conclusions, because as we will show below, even more mildly radially
anisotropic models are already strongly ruled out by the data.

Once the DF has been calculated, we are interested in the probability
${\cal F}_{\beta}(R,v) \> {\rm d}v$ that a particle observed at
projected radius $R$ is observed to have line-of-sight velocity $v$
between $v$ and $v+{\rm d}v$. This probability is given by an integral
over the DF:
\begin{equation}
  {\cal F}_{\beta}(R,v) = {1\over{\Sigma(R)}}
    \int {\rm d} z \int\!\!\!\int 
    {\rm d} v_x \> {\rm d} v_y \> f_{\beta}(E,L)  ,
\label{VPdef}
\end{equation}
where $(x,y,z)$ is a cartesian coordinate system with the $z$-axis
along the line of sight (i.e., the line-of-sight velocity $v$ equals
$v_z$). For the numerical evaluation of the outer, line-of-sight,
integral we change variables from $z$ to~$u \equiv {\rm arsinh}\>
(z/R)$.  The evaluation of the inner integral over $(v_x,v_y)$ is more
complicated. We transform to polar co-ordinates $(w,\zeta)$ defined by
$v_x \equiv w \cos \zeta$ and $v_y \equiv w \sin \zeta$. For models
with $\beta \leq 0$ we directly evaluate the resulting double integral
over $(w,\zeta)$. Models with $\beta > 0$, however, have an awkward
singularity in the DF at $L=0$. For these models we apply an
additional change of variables from $\zeta$ to $s \equiv \ln
\tan {1\over2} \zeta$.

\subsection{Data-model comparison}
\label{ss:comparison}

We used the approach of \S\ref{ss:DFcalc} to calculate the DFs and
probability distributions ${\cal F}_{\beta}(R,v)$ for each of the
eleven constant-$\beta$ models shown in Figure~\ref{f:massparam}. With
these models we study the $(R,v)$ datapoints in the CNOC1 cluster
survey for those galaxies with K-corrected absolute Gunn $r$-band
magnitude $M_r < -18.5$.  Galaxies in the clusters MS 0906+11 and MS
1358+62 were excluded; these clusters are strong binaries for which
spherical models cannot be appropriate (C97b). The sample of galaxies
from the remaining 14 survey clusters was restricted to those galaxies
with $R \leq 1.5$ and $|v| \leq v_{\rm max} \equiv 4$. So, without
loss of generality, we ignore the regions of $(R,v)$ space were
cluster members are scarce among large numbers of interlopers. The
remaining sample contains 990 galaxies.

The sample thus defined is still expected to contain interlopers.  We
have not attempted any identification and removal of individual
interlopers, but instead take interloper contamination into account in
a statistical sense. We assume: (i) that the density of interlopers at
fixed $R$ is homogeneous in $v$; and (ii) that the probability $\fint$
for a galaxy at known $R$ but unknown $v$ to be an interloper is
independent of $R$. With these assumptions, $\fint$ is the total
fraction of interlopers in the sample. While this treatment of
interlopers was motivated in part by mathematical simplicity, it does
not oversimplify things to the point where the models become
inadequate. We discuss the merits and limitations of our treatment of
interlopers in detail in \S\ref{ss:discinter} below.

\subsubsection{Likelihood analysis}
\label{sss:likeanalysis}
 
Let galaxy number $j$ in the sample be observed at radius $R_j$. The
probability ${\cal F}(R_j,v) {\rm d}v$ that the observed velocity $v$
of the galaxy falls between $v$ and $v+{\rm d}v$ is given by
\begin{equation}
  {\cal F}(R_j,v) {\rm d}v = \cases{
     \fint \> ({\rm d}v / 2 v_{\rm max}) +
     (1-\fint) \> {\hat {\cal F}}_{\beta}(R_j,v) {\rm d}v  , 
        & $|v| \leq v_{\rm max}$ ; \cr
     0 , 
        & $|v| > v_{\rm max}$ . \cr}
\label{probfunc}
\end{equation}
In this equation ${\hat {\cal F}}_{\beta}(R_j,v)$ denotes the
convolution of ${\cal F}_{\beta}(R_j,v)$ with a normalized Gaussian of
dispersion $\Delta v_j$, where $\Delta v_j$ is the formal measurement
error in the determination of the velocity $v_j$ of galaxy $j$. The
velocity errors are typically $\lta 140 \kms$, or $|\Delta v_j| \lta
0.15$ in dimensionless units (small enough to have negligible
influence on any part of our analysis). The probability ${\cal
F}(R_j,v)$ in equation~(\ref{probfunc}) is normalized to unity, since
realistic dynamical models predict ${\cal F}_{\beta}(R,v) \approx 0$
for $|v| > v_{\rm max} = 4$.

The complete dataset consists of $N$ galaxies observed at $(R_j,v_j)$,
with $j=1,\ldots,N$. The probability of this dataset in a given
cluster model is proportional to the likelihood $L \equiv
\prod_{j=1}^{N} {\cal F}(R_j,v_j)$. Instead of maximizing the
likelihood we seek to minimize the quantity\footnote{In an ideal case
in which the ${\cal F}(R_j,v)$ are all Gaussian (not true here),
$\lambda$ reduces to a $\chi^2$ statistic.}
\begin{equation}
  \lambda \equiv -2 \ln L = -2 \sum_{j=1}^{N} \ln [ {\cal F}(R_j,v_j) ] .
\label{likedef}
\end{equation}
The goal is to calculate the likelihood quantity $\lambda$ on a grid
of the model parameters $(\beta,\fint)$, and to search for the
best-fitting model that yields the minimum value $\lambda_{\rm
min}$. Two further questions then remain to be answered: (a) is the
best-fitting model statistically acceptable; and (b) what are the
confidence regions around the best-fitting model parameters? To
address the first question we resort to Monte-Carlo simulations. For
each galaxy $j$ in the sample we draw a velocity from the probability
distribution ${\cal F}(R_j,v)$ for the best-fitting model. For the
resulting pseudo-dataset we calculate the likelihood quantity
$\lambda$ as defined by equation~(\ref{likedef}). This is repeated
many times. From the resulting set of $\lambda$ values we calculate
the median, as well as the confidences region around the median that
contain a fixed fraction of the simulated $\lambda$ values (e.g.,
68.3\%, which corresponds to 1-$\sigma$ for a Gaussian distribution,
or $95.4$\% which corresponds to 2-$\sigma$). To address the second
question and obtain confidence regions around the best-fitting model
parameters, we use a well-known theorem of mathematical statistics
(e.g., Stuart \& Ord 1991; used also by Merritt \& Saha 1993): the
likelihood-ratio statistic $\lambda -
\lambda_{\rm min}$ tends to a $\chi^2$ statistic in the limit of $N
\rightarrow \infty$, with the number of degrees-of-freedom equal to
the number of free parameters that have not yet been varied and chosen
so as to optimize the fit. Hence, the likelihood-ratio statistic
$\lambda - \lambda_{\rm min}$ reduces to the well-known $\Delta
\chi^2$ statistic (e.g., Press \etal 1992) for $N \rightarrow \infty$, 
despite the fact that the ${\cal F}(R_j,v)$ are not individually
Gaussian. This is a consequence of the central limit theorem. In
principle it would be more robust to calculate the confidence regions
on the best-fitting model parameters $(\beta,\fint)$ by means of
Monte-Carlo simulation (or, e.g., bootstrapping), but in the present
context this was found to be prohibitively expensive computationally.

Figure~\ref{f:backfrac} shows the interloper fraction $\fint$ that
minimizes $\lambda$ as a function of anisotropy, with 1-$\sigma$ error
bars calculated as described above. The figure shows that $\fint$ is
in the range $\fint = 0.115 \pm 0.02$ for all anisotropies studied,
with a formal error of $\pm 0.02$ at fixed anisotropy. In the
remainder of the paper we always use, for each given anisotropy, the
optimal $\fint$ shown in Figure~\ref{f:backfrac}.
Figure~\ref{f:likebest} shows $\lambda$ as function of the velocity
dispersion anisotropy. The overall minimum is $\lambda_{\rm min} =
1783.1$ for the model with $\sigma_r / \sigma_t = 0.92$ (i.e., not far
from isotropic). Monte-Carlo drawing from this model yields a
predicted $\lambda = 1766.9 \pm 48.6$ with $68.3$\% confidence; hence,
the observed likelihood is acceptable. Confidence boundaries on the
best-fitting $\sigma_r / \sigma_t$ as obtained from the likelihood
ratio statistic $\lambda - \lambda_{\rm min}$ are indicated in the
figure. At 68.3\% confidence $0.80 \leq \sigma_r / \sigma_t \leq 1.01$
and at $95.4$\% confidence $0.48 \leq \sigma_r / \sigma_t \leq 1.11$.

The structural properties of the ensemble cluster depend on $\sigma_r
/ \sigma_t$, as shown in Figures~\ref{f:massparam}
and~\ref{f:massprof}. The inner slope $\xi$ and scale length $a$ of
the mass density, and also the mass-to-number density ratio
$\rho/\nu$, are of particular interest. Our methods do not directly
allow us to obtain confidence intervals on these quantities. However,
approximate confidence intervals are obtained by determining how
$\xi$, $a$ and $\rho/\nu$ vary in Figures~\ref{f:massparam}
and~\ref{f:massprof} when $\sigma_r / \sigma_t$ is varied over, e.g.,
its 68.3\% confidence range. This yields $0.70 \leq \xi \leq 1.13$,
$0.23 \leq a \leq 0.27$, and over the radial range $0.1
\leq R \leq 1$, $0.47 \leq \log(\rho/\nu) \leq 0.65$. The 
data are therefore consistent with an NFW profile and with a
mass-to-number density ratio that is constant to within $\sim 25$\%
over the radial range $0.1 \leq R \leq 1$, both at 68.3\% confidence.

\subsubsection{Grand-total velocity histogram}
\label{sss:histogram}
 
The likelihood is only one of many statistics that one can use to test
the validity of a model.  There are several reasons why it is useful
to quantitatively consider other statistics as well. First, other
statistics will be sensitive to different aspects of the data, and can
therefore yield different estimates of the best model and different
(possibly smaller) confidence regions. Second, there is no guarantee
that any of our models actually provides an adequate representation of
the DF that underlies the data (e.g., the ensemble cluster may not be
perfectly spherical or have non-constant $\beta$). While the
likelihood suggests that the best-fit model of
\S\ref{sss:likeanalysis} is statistically acceptable, other statistics
may well indicate that none of our models are acceptable. Third, the
likelihood analysis shows that some of our models are more likely than
others, despite the fact that all models fit the projected velocity
dispersion profile equally well. The likelihood provides little
insight into {\it why} this is the case, and other statistics may do a
better job in this respect. Based on these arguments we have
considered some other statistics that address the properties of the
grand-total velocity histogram for the ensemble cluster.

The grand-total velocity distribution predicted by a given model is
simply:
\begin{equation}
  {\cal F}_{\rm tot}(v) = {1\over N} \sum_{j=1}^{N} {\cal F}(R_j,v) ,
\label{totVP}
\end{equation}
where ${\cal F}(R_j,v)$ is given by equation~(\ref{probfunc}).
However, this method of calculating the predicted velocity
distribution for a model provides no insight into the random
fluctuations caused by the shot noise. The observed grand-total
velocity histogram is determined by a finite number of galaxies, and
this is best simulated in Monte-Carlo manner. For each model with
given anisotropy we therefore draw velocities from the probability
distributions ${\cal F}(R_j,v)$ for each of the $N$ galaxies in the
sample, which yields a simulated grand-total velocity histogram. This
procedure is repeated many times, and for each histogram bin we
calculate the median occupation of the bin, and the range around the
median that contains the occupation value for 68.3\% of the
simulations.

Figure~\ref{f:histograms} shows the predicted, normalized grand-total
velocity histograms as function of $|v|$ for three constant-$\beta$
models. In each panel the predictions are shown as a combination of
two thin lines; for each bin, the histogram occupation falls between
these lines in 68.3\% of the simulations. The thick line in each of
the panels is the observed velocity histogram. The predicted velocity
histogram is approximately Gaussian for the isotropic model; it is
more flat-topped than a Gaussian for tangentially anisotropic models,
and more centrally peaked than a Gaussian for radially anisotropic
models. These results are consistent with previous calculations (in
other contexts) of the velocity distributions predicted by constant
anisotropy models (e.g., Merritt 1987; van der Marel \& Franx 1993).
The observed histogram is neither particularly centrally peaked nor
particularly flat-topped, and this is why the likelihood analysis of
\S\ref{sss:likeanalysis} yields a best-fit model that is close to 
isotropic.

One method to address whether the differences between the observed and
predicted grand-total velocity histograms are statistically acceptable
is through a $\chi^2$ quantity that sums the squared residuals over
all bins of the velocity histogram, weighted with the shot-noise
errors predicted from the Monte-Carlo simulations. We calculated this
quantity, and found a minimum $\chi^2 = 26.1$ for the model with
$\sigma_r / \sigma_t = 0.91$. The expected value with 20 velocity bins
is $\chi^2 = 20.0 \pm 6.6$ at $68.3$\% confidence, suggesting that the
best fit model is acceptable. Confidence boundaries on $\sigma_r /
\sigma_t$ were calculated from the usual $\Delta \chi^2$ statistic,
yielding $0.74 \leq \sigma_r / \sigma_t \leq 1.02$ at 68.3\%
confidence and $0.47 \leq \sigma_r / \sigma_t \leq 1.13$ at $95.4$\%
confidence. These results are virtually identical to those derived in
\S\ref{sss:likeanalysis} from the likelihood statistic $\lambda$.
Apparently, little information is lost by putting all datapoints
together in one grand-total velocity histogram (which removes
information on radial dependencies).

More insight can be gained by considering statistics that address the
{\it shape} of the grand-total velocity histogram. By construction,
the predicted histograms for models with different $\beta$ are all
normalized, and have the same dispersion.  The most obvious shape
statistics are therefore the kurtosis and other higher order moments
(Stuart \& Ord 1991). However, these moments are very sensitive to the
wings of the velocity distribution, which are not particularly well
constrained (primarily due to interlopers). The Gauss-Hermite moments
provide more suitable statistics, since they are by construction
insensitive to the wings of the distribution (van der Marel \& Franx
1993; Gerhard 1993). These moments have been used before to describe
cluster velocity histograms, but primarily as a method for searching
for cluster substructure (Zabludoff \etal 1993; see also Colless \&
Dunn 1996).

To calculate the Gauss-Hermite moments for the CNOC1 ensemble cluster,
the observed velocities were binned into a grand-total histogram
${\cal F}_{\rm tot, observed}(v)$ as in Figure~\ref{f:histograms}. The
estimated contribution by interlopers was then subtracted to yield a
corrected histogram:
\begin{equation}
  {\cal F}_{\rm tot, corrected}(v) =
      [ {\cal F}_{\rm tot, observed}(v) -
        (\fint / 2 v_{\rm max})] \> / \> (1-\fint) ,
      \qquad |v| \leq v_{\rm max} .
\label{intersub}
\end{equation}
From this corrected histogram we calculated the two lowest order
non-trivial Gauss-Hermite moments\footnote{The best-fitting Gaussian
is used to generate the Gauss-Hermite basis, so by definition $h_0 =
1$ and $h_2 = 0$. The odd moments are all zero for a symmetrical
distribution.}, namely $h_4$ and $h_6$. The resulting moments were
found to be independent of the choice of the bin size in the histogram
construction\footnote{The underlying reason for this is that each
Gauss-Hermite moment is sensitive to features in the velocity
distribution on a particular characteristic velocity scale. This scale
becomes finer for moments of higher order, as in a Fourier
decomposition (Gerhard 1993). We find that velocity binning has no
influence on the calculated moments, as long as the bin size of the
histogram is smaller than the characteristic velocity scale for the
highest order considered, $h_6$ in this case.}; hence they are well
defined quantities. The best estimate for $\fint$ depends slightly on
the assumed anisotropy (cf.~Figure~\ref{f:backfrac}), so the inferred
$h_4$ and $h_6$ do so as well. This is seen in Figure~\ref{f:gauher},
in which dashed curves show the dependence of the observed $h_4$ and
$h_6$ on the assumed anisotropy.  However, the dependence on
anisotropy is small, and we find that $h_4$ and $h_6$ are in the range
$h_4 = -0.015 \pm 0.005$ and $h_6 = -0.028 \pm 0.006$ for all
anisotropies that we have studied.

We also calculated $h_4$ and $h_6$ for the models. As before, velocity
histograms were drawn from each model in Monte-Carlo fashion. For each
ensemble of simulated histograms we calculated the median and $68.3$\%
confidence interval on $h_4$ and $h_6$. The resulting predictions are
shown in Figure~\ref{f:backfrac} as solid dots with error bars. In
essence, the dots are the Gauss-Hermite moments predicted for the
hypothetical case in which complete information on the velocity
distribution were available at the radius of each galaxy in the
sample. The error bars show the 1-$\sigma$ shot noise variations on
these values due to the fact that only a discrete realization of the
model is available with a finite number of galaxies\footnote{In the
approach of Figure~\ref{f:gauher}, the observed Gauss-Hermite moments
are fixed numbers without error bars (they are well defined integrals
over the observed histogram), while the model predictions for the
observed values have a shot-noise error. An alternative view is to
think of the observed Gauss-Hermite moments as estimates of the
moments of the underlying distribution. In this view, the models
predict a fixed value (the dots in the figure), and the Monte-Carlo
calculated shot-noise errors should be thought of as the errors on the
observed values.}. These errors are $|\Delta h_l| \approx 0.02$--0.03,
independent of either $\beta$ or the Gauss-Hermite order
$l$\footnote{The fact that the shot-noise errors are independent of
the Gauss-Hermite order is due to the fact that the Gauss-Hermite
functions form an orthonormal basis.}.

The tangentially anisotropic models predict flat-topped velocity
histograms, which have $h_4 < 0$ and $h_6 >0$; the radially
anisotropic models predict centrally peaked velocity histograms, which
have $h_4 > 0$ and $h_6 < 0$. There is generally more power in the
fourth-order term than the sixth-order term. Even higher-order moments
are not particularly useful in the present context, since all models
that we have studied predict $h_l \approx 0$ to within the variations
expected from shot noise, for all $l \geq 8$. The near-zero observed
values of $h_4$ and $h_6$ are reproduced only by models that are close
to isotropic. The observed values fall inside the $68.3$\% confidence
interval for both $h_4$ and $h_6$ if $1.00 \leq \sigma_r / \sigma_t
\leq 1.05$. They fall inside the $95.4$\% confidence interval for both
$h_4$ and $h_6$ if $0.75 \leq \sigma_r / \sigma_t \leq 1.21$. These
results are broadly consistent with those derived in
\S\ref{sss:likeanalysis} and \S\ref{sss:histogram}, but not entirely. 
In particular, tangentially anisotropic models are ruled out at higher
confidence by $h_4$ and $h_6$ than they are by the likelihood
statistic $\lambda$.

\section{Possible uncertainties and additional considerations}
\label{ss:uncertainties}

The dynamical analysis in the preceding sections has yielded several
interesting conclusions about the structure of clusters of galaxies.
Although the analysis has been significantly more detailed than many
previous studies for this and other data sets, it still involves a
number of simplifying assumptions. The present section discusses how
the approximations in the analysis may have affected the conclusions,
and how robust the conclusions are in view of this.

\subsection{Treatment of interlopers}
\label{ss:discinter}

Our sample definition assumes that cluster members have $|v| \leq
v_{\rm max} = 4$ in units of the cluster dispersion. So we identify
all galaxies with observed velocities in excess of this limit as
interlopers, similar to the approach adopted by Yahil \& Vidal (1977)
(they use $v_{\rm max} = 3$). Other than this very conservative cut,
no removal of interlopers from the sample is attempted. This approach
differs from that used by most authors who have modeled the dynamics
of clusters of galaxies. Fairly complex schemes have recently been
developed for the identification of interlopers, with estimated
success rates of up to 90\% (e.g., Perea, del Omo \& Moles 1990; den
Hartog \& Katgert 1996). These schemes generally rely on mass
estimates obtained with specific assumptions about the orbital
structure. In the present context the orbital structure is what we
hope to determine, so use of these schemes could possibly introduce
subtle biases. Although the schemes are tailored to be robust and
conservative, we feel it is safer in the present context to model the
presence of interlopers in a statistical sense, rather than to try to
remove them. Note that this modeling would still be necessary even if
we did try to remove interlopers, because no interloper identification
scheme can be 100\% successful.

In our statistical treatment of interlopers we assumed that the
density of interlopers is homogeneous in velocity.\footnote{Note that
we are not assuming that the {\it probability} for a galaxy to be an
interloper is independent of $v$. Interlopers are assumed to have a
homogeneous density in $v$, while cluster members have a density that
increases strongly towards $|v| = 0$ (see
Figure~\ref{f:histograms}). Hence, the probability for a galaxy to be
an interloper is a strongly increasing function of $|v|$.}  This
assumption is directly verifiable, since the density of interlopers at
velocities $|v| > v_{\rm max}$ can be determined from the CNOC1
dataset. To this end we extracted from the CNOC1 survey as before the
galaxies with K-corrected absolute Gunn $r$-band magnitude $M_r <
-18.5$, with exclusion of galaxies in the clusters MS 0906+11 and MS
1358+62. From the resulting sample we calculated the density of
galaxies in $(R,v)$ space over the region defined by $R \leq 1.5$ and
$5 \leq |v| \leq 10$,\footnote{C97b also assumed the density of
interlopers to be homogeneous in velocity space (in their velocity
dispersion calculations), but they used a somewhat larger velocity
range, $5 \leq |v| \leq 25$, to determine the interloper density. This
yields results that do not differ significantly from those obtained
here.}  and also over the region defined by $R \leq 1.5$ and $|v| \leq
v_{\rm max} = 4$. Comparison of these densities yields a direct
estimate of the interloper fraction $\fint$ for the sample studied in
\S\ref{s:DFmodels}. This approach yields $\fint = 0.117$, fully consistent 
with the values inferred from the likelihood analysis shown in
Figure~\ref{f:backfrac} (the best-fitting model identified by
Figure~\ref{f:likebest} has $\fint = 0.108 \pm 0.017$). This provides
direct evidence that the density of interlopers is similar for $|v|
\leq v_{\rm max} = 4$ as for $5 \leq |v| \leq 10$, thus supporting the
assumption underlying our analysis. This should also serve as a
warning for interloper removal schemes, which can never identify
interlopers with small velocities.

The second assumption in our analysis is that the probability $\fint$
for a galaxy at known $R$ but unknown $v$ to be an interloper is
independent of $R$. This appears somewhat counter-intuitive, since the
surface density $\Sigma(R)$ of cluster members is strongly peaked
towards $R=0$. So if the surface density of interlopers $\Sigma_{\rm
int}(R)$ is homogeneous, one would expect $\fint$ to decrease towards
$R=0$ approximately as $\fint \propto [\Sigma(R)]^{-1}$. By contrast,
our assumption implies $\Sigma_{\rm int}(R) \propto \Sigma(R)$, i.e.,
that the density of interlopers (with $|v| \leq v_{\rm max} = 4$)
increases as steeply towards $R=0$ as does the density of cluster
members. This may not necessarily be incorrect, because matter not
belonging to the cluster may in fact be strongly clustered towards
it. To address the validity of our assumption we have taken an
empirical approach, by dividing the sample of \S\ref{s:DFmodels} in
two equally sized subsamples of galaxies at small and large $R$,
respectively.  The first subsample contains galaxies with $R \leq
0.41$, and has a median radius $R_{{\rm med},1} = 0.23$; the second
subsample contains galaxies with $ 0.42 \leq R \leq 1.5$, and has a
median radius $R_{{\rm med},2} = 0.73$. For each subsample we redid
the likelihood analysis, using an isotropic model. This yields
interloper fractions $f_{\rm i,1} = 0.083 \pm 0.024$ for the first
subsample, and $f_{\rm i,2} = 0.125 \pm 0.025$ for the second
subsample. So $\fint$ does appear to decrease with decreasing radius,
but this is barely significant at the 1-$\sigma$ level. However,
$\Sigma(R_{{\rm med},1})$ is $\sim 5$ times as large as
$\Sigma(R_{{\rm med},2})$, so $\fint$ certainly does not decrease
towards $R=0$ as fast as predicted if $\Sigma_{\rm int}(R)$ were
independent of $R$. So although our assumption may not be fully
correct, it does seem acceptable.

All in all, none of the tests that we have done has given us reason to
believe that our treatment of interlopers is significantly in error.
Also, interlopers make up only a small fraction ($\sim 11$\%) of the
sample studied in \S\ref{s:DFmodels}, and we do not expect the final
results to be particularly sensitive to their treatment.

\subsection{Choice of distribution functions}
\label{ss:discDF}
 
There are several restrictions to the dynamical analysis that we have
employed. Although we construct phase-space distribution functions and
model the full CNOC1 $(R,v)$ dataset, we do restrict ourselves to a
specific set of models. We study only models with constant anisotropy,
and even among the DFs that generate models with constant anisotropy,
we choose a particular set. Realistic systems are unlikely to have a
velocity dispersion anisotropy that is independent of radius, and it
is not a priori clear whether our results can provide any information
on what range of such models could be acceptable.

Despite the shortcomings of our analysis, there are reasons to believe
that our results may be more robust than they would otherwise seem. In
a nutshell, our analysis boils down to the fact that the observed
grand-total velocity histogram for the CNOC1 sample is close to
Gaussian, and this is not generally expected for models that have
significant velocity anisotropy. Van der Marel \& Franx (1993)
presented a simple method for calculating the Gauss-Hermite moments
for constant-$\beta$ models with various scale-free number densities
and potentials. Such models show that the Gauss-Hermite moments are
determined primarily by the overall velocity dispersion anisotropy,
and to much smaller extent by the details of the number density
profile and potential. Gerhard (1993) reached the same conclusion for
models with varying $\beta$. So even though we have parameterized
$\nu(r)$, $\rho(r)$ and $\beta$ in our approach, it is likely that
even with a more general non-parametric approach the observations
would still imply a velocity distribution that is close to isotropic.

The approach that we have used shares many similarities to that used
by Merritt \& Saha (1993) to interpret 296 measured velocities for the
Coma cluster. In particular, we follow their approach of maximizing
the likelihood for the $(R,v)$ dataset. Like us, they use a restricted
parameterization for the mass profile; while they adopt a
two-parameter family for the gravitational potential, we adopt a
three-parameter family for the mass density. The technique of Merritt
\& Saha does have two important advantages over our approach: first, 
it makes a basis function expansion of the DF that allows fairly
arbitrary anisotropy profiles; and second, it avoids binning and
parameterization of the projected number density profile. On the other
hand, our method expands on that of Merritt \& Saha through the
inclusion of explicit modeling of interloper contamination, the
Monte-Carlo simulation of the dataset to estimate confidence regions
and the goodness-of-fit, and the use of Gauss-Hermite moments. A
hybrid version of our techniques would probably allow the most robust
conclusions to be reached, but this is beyond the scope of the present
paper.

\subsection{Brightest Cluster Galaxies}
\label{ss:discBCG}

Our analysis has included the brightest cluster galaxies (BCGs) of
each of the clusters in the sample. BCGs have very special properties;
they tend to be atypically bright, and they are generally found at
$(R,v) \approx (0,0)$. So it is not clear whether it is appropriate to
treat them on a par with the other cluster members, as we have
done. Exclusion of the BCGs from our sample would, among other things,
decrease the projected number density $\Sigma(R)$ at small radii.
Specification of $\Sigma(R)$ is the first step in our modeling
approach, so our entire analysis would need to be repeated to
determine accurately whether exclusion of the BCG's would alter the
results of our dynamical models. To avoid this, we have chosen a more
approximate route to address this issue. We removed the BCG's from the
sample, and then recalculated the Gauss-Hermite moments of the
observed grand-total velocity histogram. The BCGs tend to have $v
\approx 0$, so their removal makes the histogram more flat-topped. We
find that $h_4$ and $h_6$ for the resulting sample are in the range
$h_4 = -0.024 \pm 0.005$ and $h_6 = -0.022 \pm 0.006$ independent of
the assumed anisotropy (which, as in Figure~\ref{f:gauher}, influences
the Gauss-Hermite moments through the best-fitting interloper fraction
$\fint$). As discussed in \S\ref{ss:discDF}, the Gauss-Hermite
moments predicted by dynamical models depend primarily on the
anisotropy, and to a much lesser extent on the number density profile.
So it may be reasonable to use the theoretical relation between $h_4$,
$h_6$ and anisotropy in Figure~\ref{f:gauher}, despite the fact that
it was derived for a somewhat different $\Sigma(R)$. The observed
estimates of $h_4$ and $h_6$ without BCGs then imply best-fitting
models that are somewhat more tangentially anisotropic than those
derived in \S\ref{s:DFmodels}, but only by $\Delta
\log(\sigma_r/\sigma_t) = -0.03$.  This change is small compared to
the size of the confidence regions on the best-fitting models. Hence,
the inclusion or removal of BCGs has no significant effect on the main
conclusions from our analysis.

\subsection{Non-sphericity}
\label{ss:axisym}

Clusters of galaxies are generally flattened. The mode of the
distribution of projected axial ratios for clusters is $q \approx
0.6$, while the mode of the distribution of intrinsic axial ratios is
$q \approx 0.45$ (de Theije, Katgert, \& van Kampen 1995). It is
important to know what this implies for the validity of our spherical
modeling.

\subsubsection{Grand-total velocity distribution}
\label{ss:axisymGT}

We consider as a simple test case axisymmetric clusters of fixed
intrinsic axial ratio $q$, with number density profiles of the form
studied by Hernquist (1990),
\begin{equation}
  \nu(R,z) = {1 \over {2\pi q}} \> m^{-1}\ (1+m)^{-3} , \qquad 
  m^2 \equiv R^2 + (z^2/q^2) ,
\label{Hernqdef}
\end{equation}
and constant mass-to-number density ratios $\rho/\nu = 1$. These
models have scale length $a=1$ and total mass $M=1$. We restrict
ourself to the simplest type of axisymmetric systems, namely those in
which the DF depends only on the two classical integrals of motion, $f
= f(E,L_z)$. We denote by ${\cal F}_{q}(R,\eta; v; i) \> {\rm d}v$ the
probability that a star at position $(R,\eta)$ in a polar coordinate
system on the sky, residing in a system of axial ratio $q$ that is
viewed at inclination $i$, is observed to have line-of-sight velocity
$v$ between $v$ and $v+{\rm d}v$. The probability distributions ${\cal
F}_{q}$ can be conveniently calculated for two-integral models using,
e.g., the method of Magorrian \& Binney (1994). The grand-total
velocity probability distribution for a system of given $q$ and $i$ is
given by
\begin{equation}
  {\cal F}_{q}(v; i) = 
     \int \Sigma_{q}(R,\eta;i) \> {\cal F}_{q}(R,\eta; v; i) \> 
          R \> {\rm d}R \> {\rm d}\eta \quad \Big / \quad 
     \int \Sigma_{q}(R,\eta;i) \> R \> {\rm d}R \> {\rm d}\eta , 
\label{skyints}
\end{equation}
where $\Sigma_{q}(R,\eta;i)$ is the projected number density obtained
when the $\nu(R,z)$ given by equation~(\ref{Hernqdef}) is viewed at
the given inclination, and where the integrals extend over the
two-dimensional plane of the sky. We denote by $\sigma_{q}(i)$ the
velocity dispersion corresponding to the distribution ${\cal F}_{q}(v;
i)$. We consider now the case of a large number of identical
axisymmetric clusters of fixed intrinsic axial ratio $q$, viewed from
random viewing directions. From these clusters we build an ensemble
cluster, as we have done for the CNOC1 data set. The grand-total
velocity probability distribution of the ensemble cluster is given by
\begin{equation}
  {\cal F}_{q}({\tilde v}) =
     \int [{\cal F}_{q}(v/\sigma_{q}(i); i) / \sigma_{q}(i)] \>
          \sin i \> {\rm d}i \quad \Big / \quad
     \int \sin i \> {\rm d}i ,
\label{inclint}
\end{equation}
where the integral is over all inclination angles $i \in [0,\pi/2]$,
and where ${\tilde v}$ is now a dimensionless velocity.

The velocity anisotropy of an $f(E,L_z)$ model with fixed axial ratio
$q$ can be characterized by the quantity
\begin{equation}
   \langle \sigma_r / \sigma_t \rangle \> \equiv \>
      \left [ \> 
         2 \langle v_r^2 \rangle \> / \> 
         ( \langle v_{\phi}^2 \rangle + \langle v_{\theta}^2 \rangle ) 
      \> \right ]^{1/2} ,
\label{anisomassdef}
\end{equation}
where the angle brackets denote mass-weighted averages over the
system. For a spherical system with constant anisotropy, $\langle
\sigma_r / \sigma_t \rangle$ reduces to the same quantity 
$\sigma_r / \sigma_t$ that we have used to characterize anisotropy in,
e.g., Figures~\ref{f:massparam}, \ref{f:backfrac},
\ref{f:likebest} and~\ref{f:gauher}. For an axisymmetric $f = f(E,L_z)$
system, $\langle \sigma_r / \sigma_t \rangle$ is determined uniquely
by the axial ratio $q$, according to the tensor virial theorem (Binney
\& Tremaine 1987). For an oblate two-integral system,  
\begin{equation}
   \langle \sigma_r / \sigma_t \rangle_q =
    \left [ {
    { \Bigl ( 1 - \sqrt{1 - e^2} {{\arcsin e} \over e}
      \Bigr ) } \> \Big / \>
    { \Bigl ( {1 \over {\sqrt{1 - e^2}}}
                    {{\arcsin e} \over e} - 1 \Bigr ) } }
    \> \right ]^{1/2} ,
\end{equation}
where the eccentricity is defined by $e^2 \equiv 1 - q^2$ .

For several values of the axial ratio ranging from $q=0.3$ to $q=1$ we
calculated the velocity distributions ${\cal F}_{q}({\tilde v})$ as
pertaining to an ensemble of randomly oriented clusters. In the
integrals of equation~(\ref{skyints}) we included only radii $R \in
[0.1 ; 3]$ to approximate the range of radii for which data is
available in the CNOC1 data set. From each distribution we calculated
the corresponding fourth-order Gauss-Hermite moments $h_{4,q}$, which
describe the extent to which the distributions deviate from Gaussians,
and for each axial ratio we also calculated the velocity anisotropy
parameter $\langle \sigma_r / \sigma_t \rangle_q$. The dots in
Figure~\ref{f:axiherm} connected by a solid curve show $h_{4,q}$ as
function of $\langle \sigma_r / \sigma_t \rangle_q$. For comparison,
we also calculated the relation between $h_4$ and $\sigma_r /
\sigma_t$ for {\it spherical} constant-$\beta$ models with a Hernquist
density profile, using the approach of \S\ref{ss:DFcalc}. Those
results are shown as a dashed curve in Figure~\ref{f:axiherm}. The
results of these two very different calculations are virtually
identical.

\subsubsection{Radial dependence of model quantities}
\label{ss:axisymRAD}

Figure~\ref{f:axiherm} suggests that it is reasonable to model the
grand-total velocity distribution of an ensemble of axisymmetric
clusters with a spherical model. However, for a spherical model to be
appropriate it must also predict the correct radial dependence for all
observable quantities. The projected intensity profile of an ensemble
of axisymmetric clusters is given by
\begin{equation}
  \Sigma_{q}(R) = {1\over{2\pi}} \> \int \Sigma_{q}(R,\eta;i) \> 
                  {\rm d}\eta \> \sin i \> {\rm d}i ,
\label{ensembintprof}
\end{equation}
where the integrals are over $\eta \in [0,2\pi]$ and $i \in
[0,\pi/2]$. The radial profiles of the kinematical quantities are
obtained by evaluating equation~(\ref{skyints}) without the integrals
over $R \> {\rm d}R$, followed by integration over inclinations as in
equation~(\ref{inclint}). This yields velocity probability
distributions from which one can calculate the run of the velocity
dispersion and the Gauss-Hermite moment $h_4$. We did these
calculations for an ensemble of two-integral Hernquist models with
axial ratio $q=0.6$; the solid curves in Figure~\ref{f:axiRfunc} show
the resulting profiles.

We used the approach of \S\ref{ss:DFcalc} to calculate a
constant-$\beta$ DF for the spherical model that has the same
projected intensity profile as the ensemble of axisymmetric clusters,
with $\sigma_r / \sigma_t$ fixed to the value $\langle
\sigma_r / \sigma_t \rangle = 0.81$ that applies to an $f(E,L_z)$
model with $q=0.6$. The dashed curves in the bottom two panels of
Figure~\ref{f:axiRfunc} show the resulting $\sigma(R)$ and
$h_4(R)$. The profiles are closely similar to those for the ensemble
cluster, with residuals $| \Delta \log \sigma |
\lta 0.04$ and $| \Delta h_4 | \lta 0.02$ at all radii of interest.
The projected intensity profile of the ensemble cluster is almost
identical (residuals $| \Delta \Sigma | \lta 0.01$) to the projected
intensity profile of a spherical Hernquist model with a scale length
$a$ for which $\log a = -0.046$. This also differs only slightly from
the value ($\log a = 0$) for the axisymmetric clusters from which the
ensemble was constructed. 

The curves for the ensemble in Figure~\ref{f:axiRfunc} were obtained
without scaling each cluster with its individual velocity dispersion
$\sigma_{q}(i)$ in equation~(\ref{inclint}). The normalizations of
$\sigma(R)$ for the ensemble cluster and the spherical model are
therefore similar in an absolute sense.\footnote{It was verified that
the agreement between $\sigma(R)$ and $h_4(R)$ for the ensemble
cluster and the spherical model does not deteriorate when one does
include the scaling of clusters with their individual velocity
dispersion (as we have done for the CNOC1 dataset).} This is important
for cluster mass determinations, since $M \propto \sigma^2$. It
implies that the average mass (or mass-to-light ratio) of a set of
clusters can be adequately determined from spherical models, even if
the clusters are individually axisymmetric.

\subsubsection{Dependence on axial ratio and anisotropy}
\label{ss:axisymASYMP}

The $f(E,L_z)$ models discussed so far provide only one possible
dynamical structure for axisymmetric systems. In general, the
construction of DFs for axisymmetric systems is a complicated problem
(e.g., van der Marel \etal 1998; Cretton \etal 1999) that is beyond
the scope of the present paper. However, if we restrict ourselves to
the case of asymptotically large radii then the problem becomes more
tractable. An axisymmetric Hernquist model then reduces to a
scale-free spheroidal mass density in a Kepler potential. Two families
of three-integral DFs for this limiting case were presented by de
Bruijne \etal (1996). Each family has one free parameter called
$\beta$. In the spherical limit both families reduce to the spherical
constant-$\beta$ models that we have discussed in \S\ref{ss:DFcalc},
but in the more general axisymmetric case the families have different
properties. The `case~I' DFs for a given axial ratio are
generalizations of the $f(E,L_z)$ models; they reduce to the
$f(E,L_z)$ model for $\beta=0$. The `case~II' DFs for a given axial
ratio correspond to models in which the velocity anisotropy $\beta
\equiv 1 - \sigma^2_t/\sigma^2_r$ is constant throughout the system,
whereas the anisotropies $\sigma_{\theta} / \sigma_r$ and
$\sigma_{\phi} / \sigma_r$ are not.  The case~II DFs have a
particularly interesting property in the present context. If one
constructs an ensemble of axisymmetric models with this DF, then the
entire projected velocity distribution of this ensemble (i.e., the
velocity dispersion and all Gauss-Hermite moments) is identical to
that for a spherical model with the same $\beta$. This is true for any
axial ratio $q$ and velocity anisotropy $\beta$. So if clusters of
galaxies are axisymmetric with DFs of this type, then the construction
of spherical models for an ensemble of such clusters will yield {\it
exactly} the correct results (at large radii), even for extreme axial
ratios or anisotropies.

In general, the differences between the predictions of an ensemble of
axisymmetric systems and a corresponding spherical model will depend
on the details of the shapes and DFs of clusters of galaxies, which
are not known. However, the combined results of all the tests that we
have done indicate that the agreement in all quantities of interest is
generally quite good. Hence, it appears adequate to use spherical
models to interpret data for an ensemble of clusters that may
individually not be spherical.

\subsection{Homology}
\label{ss:homology}

The CNOC1 ensemble cluster that we have analyzed was constructed under
the assumption that clusters form a homologous set. This allowed us to
scale all data in both radius and velocity. The virial theorem for a
collisionless system states that $\sigma^2 = GM / r_{\rm g}$, where
$\sigma$ is the grand-total velocity dispersion of the system, $M$ is
its mass, and $r_{\rm g}$ is the gravitational radius (Binney \&
Tremaine 1987). To properly scale the data for a homologous set of
clusters, one would need to scale all velocities by $\sigma$ and all
radii by $r_{\rm g}$. However, $r_{\rm g} \equiv GM^2 / W$ is defined
in terms of the potential energy $W$ of the system, which itself
depends on the exact radial profile of the mass density (Binney \&
Tremaine 1987). This is what we wish to determine, and $r_{\rm g}$ is
therefore not known a priori for individual clusters. Lacking
knowledge of $r_{\rm g}$, we scaled all radii with $r_{200}$ when
building the CNOC1 ensemble cluster in
\S\ref{s:jeansmodels}.

One may wonder whether the scaling of the data with $r_{200}$ may have
biased the results of our analysis in some way. The likely answer is
that it hasn't. In \S\ref{sss:histogram} we compared model predictions
to the grand-total CNOC1 velocity histogram. This histogram is
entirely independent of the radial scaling adopted in the construction
of the ensemble cluster; it only depends on the velocity
scaling. Nonetheless, the results obtained from the grand-total CNOC1
velocity histogram are in good agreement with those obtained from the
more complete likelihood analysis, which does depend on the details of
the radial scaling adopted in the building of the ensemble
cluster. This indicates that our results are robust, and not
particularly sensitive to the adopted radial scaling.

The radii $r_{200}$ for the CNOC1 clusters were calculated from the
relation $r_{200} = \sqrt{3} \sigma / 10 H(z)$, where $H(z)$ is the
Hubble constant at redshift $z$ (C97b). To assess the influence of the
radial scaling further we did some tests with ensembles of
hypothetical clusters that obey a relation of the form $R \propto
\sigma^{\tau}$ with $\tau \not= 1$. These were then analyzed under the 
assumption $R \propto \sigma$ that was made in the analysis of the
CNOC1 data. These tests confirmed that the results of our analysis are
not very sensitive to possible errors in the adopted radial scaling,
for reasonable values of $\tau$ (Section~3.1 of Schaeffer
\etal 1993 indicates $\tau \approx 1.15 \pm 0.3$, so the scaling adopted 
for the CNOC1 dataset is itself not inconsistent with the data).

There is some evidence for the fact that clusters may form a
homologous set from the fact that their global properties appear to
lie on a fundamental plane, as for elliptical galaxies (Schaeffer
\etal 1993). However, it is still quite possible that clusters are
in fact not homologous at all. If so, our results would only (at best)
be valid in some average sense. For example, our conclusion that the
CNOC1 ensemble cluster is consistent with an isotropic velocity
distribution may well be consistent with the (contrived) hypothesis
that half of the CNOC1 clusters are strongly radially anisotropic and
the other half are strongly tangentially anisotropic.

The possible presence of cluster substructure may be another cause of
non-homology. However, the two clusters in which substructure is most
readily apparent, MS 0906+11 and MS 1358+62, were excluded from the
analysis. X-ray images for the remaining clusters appear mostly
regular, and dynamical and X-ray mass estimates agree well (Lewis
\etal 1999); this argues against significant substructure. Also, 
the procedure of adding individual clusters together minimizes the
effect of any possible substructure in the dynamical analysis.

\section{Discussion and conclusions}
\label{s:discconc}

In the context of the CNOC1 cluster survey, redshifts were obtained
for galaxies in 16 clusters at $z = 0.17$--$0.55$, selected on the
basis of their X-ray luminosity (e.g., C96). The resulting sample is
ideally suited for an analysis of the internal velocity and mass
distribution of clusters of galaxies. Previous analyses of this
dataset were based on Jeans equation models for the projected velocity
dispersion profile $\sigma(R)$ (C97b,c). The results of such models
always have a strong degeneracy between the mass density profile
$\rho(r)$ and the velocity dispersion anisotropy profile $\beta(r)$,
because these two functions of one variable are only constrained by
one function of one variable. Here we have attempted to break this
degeneracy by analyzing not only $\sigma(R)$, but the full $(R,v)$
dataset of the CNOC1 cluster survey. The sample consists of 990
galaxies with $R \leq 1.5$ and $|v| \leq v_{\rm max} = 4$ in
dimensionless units.

In \S\ref{s:jeansmodels} we repeated some of the previous Jeans
modeling, although with a somewhat different approach from that
employed by C97b,c. The results provide an illustration of the
degeneracies involved in the modeling, and serve as a useful starting
point for a more detailed analysis. In \S\ref{s:DFmodels} we presented
an analysis of the full $(R,v)$ dataset, using a one-parameter family
of models with different constant velocity dispersion anisotropy, each
using a different mass density profile $\rho(r)$ and providing the
same acceptable fit to $\sigma(R)$. The best-fit model was sought
using a variety of statistics, including the likelihood of the
dataset, and the Gauss-Hermite moments of the grand-total velocity
histogram. The confidence regions and goodness-of-fit for the best-fit
model were determined using Monte-Carlo simulations. Although the
results differ slightly depending on which statistic is used, all
statistics agree that best-fit model is close to isotropic. The
isotropic model is acceptable at the 1-$\sigma$ confidence level for
all statistics used. For none of the statistics does the 1-$\sigma$
confidence region extend below $\sigma_r / \sigma_t = 0.74$, or above
$\sigma_r / \sigma_t = 1.05$.

Cosmological $N$-body simulations for galaxy clusters generally
predict velocity distributions that are isotropic near the center, and
that become somewhat radially anisotropic towards $\sim r_{200}$
(e.g., Crone \etal 1994; Cole \& Lacey 1996; Ghigna \etal 1998). The
maximum radial anisotropy is not large, $\sigma_r / \sigma_t \approx
1.3$. The number density weighted anisotropy over the region sampled
by the CNOC1 data is $\sigma_r / \sigma_t \approx 1.1$. Although this
value is only allowed by our analysis at the 2-$\sigma$ level, theory
and observations do seem to agree that the velocity distribution of
galaxy clusters is not strongly anisotropic.

NFW have argued that dark matter halos have a universal mass density
profile characterized by three main parameters: an inner power-law
slope $\xi = 1$; an outer power-law slope $3$; and a characteristic
scale $a$ where the profile changes from its inner to its outer slope,
$a \approx 0.2$--$0.3$ in units of $r_{200}$ (depending somewhat on
the adopted cosmology; cf.~C97c). In our models we have parameterized
the cluster mass density similarly as NFW, but with $\xi$ and $a$ as
free parameters. Models with different velocity dispersion anisotropy
require very different values of $\xi$ and $a$ to fit the projected
velocity dispersion profile. Only models that are close to isotropic
have $\xi \approx 1$ (cf.~Figure~\ref{f:massparam}), and our analysis
of the full $(R,v)$ dataset shows that such models are in fact the
only ones that provide a statistically acceptable fit to the
data. Such models have $a \approx 0.24$, consistent with the
predictions of NFW. For these models the logarithmic slope of
$\rho(r)$ at the last data point ($R = 1.5$) is $\sim 2.7$, so it
cannot be established whether the mass density slope actually
converges to $3$ at large radii, as predicted by NFW. These models
have an approximately constant mass-to-number density ratio, by
contrast to the strongly anisotropic models that are inconsistent with
the data (cf.~Figure~\ref{f:massparam}).

In \S\ref{ss:uncertainties} we have discussed a number of possible
uncertainties in our analysis, including our treatment of interlopers
and BCGs, our use of a restricted one-parameter family of distribution
functions, our use of spherical models for what is in reality an
ensemble of non-spherical clusters, and our assumption that clusters
form a homologous set. The discussions and tests that we have
presented on these issues have not provided us with any serious
reasons to mistrust our results. Nonetheless, it remains true that
there are a number of important approximations in our treatment. Until
more detailed models are constructed, it will remain difficult to
fully assess all implications of the assumptions in our
analysis. Possibly the most important caveat in our results remains
the fact that we have studied an ensemble cluster, built by co-adding
data from 14 individual clusters under the assumption of
homology. This has the advantage of reducing the influence of
substructure and non-sphericity in individual clusters on the final
dataset, but if clusters are not a homologous set, then our results
will (at best) only be valid in an average sense.

To conclude, we have presented evidence that suggests that clusters of
galaxies have approximately: (i) an isotropic velocity distribution;
(ii) a mass density profile as predicted by NFW; and (iii) a
mass-to-number density ratio that is constant with radius. At the very
least, it has been shown that these properties are not inconsistent
with the CNOC1 survey data. Additional weight is added to these
conclusions by the fact that they are consistent with the preliminary
results from a weak-lensing study of a set of 10 galaxy clusters with
the Big Throughput Camera at CTIO. That study also yields mass density
profiles that are well fit by the NFW parameterization and
mass-to-light ratios that are approximately constant with radius
(Dell'Antonio \etal 2000).


\acknowledgments


\clearpage
\appendix

\section{The accuracy of simple kinematic models for cluster velocity 
histograms}

Ramirez \& De Souza (1998) and Ramirez, de Souza \& Schade (1999)
recently presented an alternative method to model cluster velocity
histograms. They studied a total of 21 clusters (including nine that
were observed in the context of the CNOC1 survey) to constrain the
orbital anisotropy of galaxies of different morphological types. The
deviations of the cluster velocity histograms from Gaussians were
quantified using the higher-order statistic $|u| \equiv (1/N)
\sum_{j=1}^{N} |v_j / \sigma|$ (this can be viewed as an alternative
to using the Gauss-Hermite moments or kurtosis). To interpret the
observed values of $|u|$ they were compared to the predictions of a
simple kinematical model that assumes that: (i) the velocity
dispersions $\sigma_r$, $\sigma_{\theta}$ and $\sigma_{\phi}$ are
constant through the system; and (ii) the velocity distribution is a
(three-dimensional) Gaussian at any point in the system. These
assumptions remove the laws of gravitational dynamics (both Poisson's
equation and the collisionless Boltzmann equation) from the problem,
as well as any information contained in (and dependence on) the galaxy
number density profile $\nu(r)$, the mass-density profile $\rho(r)$
and the projected velocity dispersion profile $\sigma(R)$. Instead,
the shape of the cluster velocity histogram in these models is a
simple unique function of only one parameter, $\sigma_r/\sigma_t$. The
assumptions on which these models are based are not generally correct,
but Ramirez \etal argue that they are reasonable and adequate for the
problem at hand. As we will show, this is not actually the case.

The primary shortcoming of the Ramirez \etal models is the assumption
of locally Gaussian velocity distributions. This is most easily seen
for the case of very tangentially anisotropic models, which have large
numbers of galaxies on (nearly) circular orbits. This leads to local
velocity distributions that are strongly bimodal, with peaks at plus
and minus the local circular velocity. Even after projection, such
models often predict velocity distributions with pronounced double
peaks (e.g., van der Marel \& Franx 1993). This behavior of real
dynamical systems is not reproduced by kinematical models that assume
locally Gaussian velocity distributions. In such models the
line-of-sight velocity distribution is unimodal everywhere along the
line of sight, and hence the same is always true for the projected
velocity distribution.

Calculations confirm these arguments and quantify the size of the
errors that are introduced. As an example, we considered models for
the CNOC1 ensemble cluster in which the velocity dispersion profiles
$\sigma_r(r)$ and $\sigma_t(r)$ were calculated from the Jeans
equations (as in \S\ref{s:jeansmodels}), but in which the velocity
distribution along the line-of-sight at any point in the system was
incorrectly assumed to be a Gaussian (with dispersion given by the
Jeans equations), instead of being calculated from a DF (as in
\S\ref{s:DFmodels}). The predicted grand-total velocity histograms
were calculated for different values of $\sigma_r/\sigma_t$, and their
shapes were quantified through the Gauss-Hermite moments $h_4$ and
$h_6$ (dashed curves in Figure~\ref{f:gaussassump}). The results can
be compared to those obtained with a self-consistent dynamical model
based on a DF (solid curves in the same figure; same as the
predictions in Figure~\ref{f:gauher}).  It is clear that the results
obtained with the simplified analysis have little in common with the
self-consistent results. The former always yields histograms that are
more centrally peaked than a Gaussian ($h_4 > 0$), and only isotropic
models produce histograms that are close to Gaussian ($h_4
\approx 0$). Ramirez \etal find the same general behavior. By
contrast, self-consistent dynamical models predict a smooth transition
from flat-topped to centrally-peaked profiles when going from
tangential anisotropy to radial anisotropy. So while the data
presented by Ramirez \etal on the velocity histogram shapes for
galaxies of different morphological types are very interesting and
definitely worth further study, it appears that their kinematical
models are insufficient to obtain reliable conclusions about orbital
anisotropies.

\clearpage


\ifsubmode\else
\baselineskip=10pt
\fi


\clearpage


\ifsubmode\else
\baselineskip=14pt
\fi


\newcommand{\figcapprojnumden}{The projected galaxy number density profile 
$\Sigma(R)$ of the CNOC1 ensemble cluster as derived in C97b. The curve
shows the parameterized fit given by equation~(\ref{projnumden}),
which is used in the dynamical modeling. The units along the axes are
dimensionless, as described in the text.\label{f:projnumden}}

\newcommand{\figcapdispprof}{The projected velocity dispersion profile
$\sigma(R)$ of the CNOC1 ensemble cluster. A 101-point running average
is shown, sampled at intervals of $0.1$ in the dimensionless projected
radius $R$. The curves show the predictions of eleven models, each
with different constant velocity dispersion anisotropy $\beta$. The
models all provide similar, statistically acceptable fits to the data,
but each uses a different mass density profile
$\rho(r)$.\label{f:dispprof}}

\newcommand{\figcapmassparam}{Parameters of the three-dimensional mass 
density $\rho(r)$ displayed in Figure~\ref{f:massprof}, as function of
the velocity dispersion anisotropy $\sigma_r / \sigma_t$, with from
top to bottom: the scale density $\rho_0$; the scale radius $a$; and
the central power-law slope $\xi$. Solid dots indicate the values of
$\sigma_r / \sigma_t$ for which detailed models are constructed in
this paper.\label{f:massparam}}

\newcommand{\figcapmassprof}{Inferred structure of the CNOC1 ensemble cluster,
with from top left to bottom right: the three-dimensional galaxy
galaxy number density $\nu(r)$; the three-dimensional mass density
$\rho(r)$; the mass-to-number-density ratio $\rho/\nu(r)$; and the
enclosed mass $M(r)$. The galaxy number density $\nu(r)$ is obtained
by deprojection of the projected galaxy number density $\Sigma(R)$
shown in Figure~\ref{f:projnumden}. The mass density $\rho(r)$ and
enclosed mass $M(r)$ are obtained by solving the Jeans equation for a
spherical system so as to best fit the projected velocity dispersion
profile shown in Figure~\ref{f:dispprof}. Eleven curves are shown,
each indicating the best fit for a different constant velocity
dispersion anisotropy $\beta$. The $\beta$-values of the models are
logarithmically spaced in $\sigma_r / \sigma_t$.  Heavy curves are for
the best-fitting isotropic model, short dashed curves are for the
best-fitting model with $\sigma_r / \sigma_t = 1/3$, and long dashed
curves are for the best-fitting model with $\sigma_r /
\sigma_t = 3$. The radial range shown along the abscissa corresponds 
approximately to the range for which the models are meaningfully
constrained by the kinematical data.\label{f:massprof}}

\newcommand{\figcapbackfrac}{Inferred fraction $\fint$ of interloper galaxies 
in the sample with 1-$\sigma$ error bars, as function of the velocity
dispersion anisotropy $\sigma_r / \sigma_t$.\label{f:backfrac}}

\newcommand{\figcaplikebest}{The likelihood quantity $\lambda$ defined in 
equation~(\ref{likedef}) as function of the velocity dispersion
anisotropy $\sigma_r / \sigma_t$. Solid points show models that were
calculated; the curve is a spline fit through the points. The best fit
model has $\sigma_r / \sigma_t = 0.92$, and is close to isotropic.
The 68.3\% and $95.4$\% confidence boundaries are indicated, as
inferred from the likelihood-ratio statistic $\lambda-\lambda_{\rm
min}$.\label{f:likebest}}

\newcommand{\figcaphistograms}{The heavy curve in each panel is the
normalized grand-total velocity histogram for the CNOC1 ensemble
cluster, as function of $|v|$. The regions between the two thin curves
are the predictions of models with, from left to right, $\sigma_r /
\sigma_t = 0.52$, $1.00$ and $1.93$. The model predictions take into
account the shot noise due to the finite number of galaxies. In
Monte-Carlo simulations the occupancy in each bin falls between the
two thin curves in 68.3\% of the drawings. Tangentially anisotropic
models yield histograms that are more flat-topped than a Gaussian
while radially anisotropic models yield histograms that are more
centrally peaked. Of the models that are displayed, the isotropic
model (middle panel) provides the best fit to the
data.\label{f:histograms}}

\newcommand{\figcapgauher}{The Gauss-Hermite moments $h_4$ and $h_6$ of the
grand-total velocity histogram for the CNOC1 ensemble cluster, as
function of the velocity dispersion anisotropy $\sigma_r /
\sigma_t$. The contribution from interlopers was subtracted as 
described in the text. The dashed curves indicate the values
calculated for the observed dataset; these values depend mildly on the
assumed anisotropy, because the best estimate for the fraction of
interloper galaxies in the sample does. The solid points indicate the
predictions of the dynamical models; the error bars are the $68.3$\%
confidence regions calculated from Monte-Carlo simulations which take
into account the finite number of galaxies in the sample.  Only models
that are close to isotropic provide an acceptable fit to both $h_4$
and $h_6$.\label{f:gauher}}

\newcommand{\figcapaxiherm}{Dots connected by a solid curve are the 
properties calculated for an ensemble of axisymmetric two-integral
Hernquist models of fixed axial ratio $q$, seen from random viewing
directions. The Gauss-Hermite moment $h_4$ refers to the grand-total
ensemble velocity histogram, and $\langle \sigma_r / \sigma_t \rangle$
is the velocity anisotropy defined in equation~(\ref{anisomassdef}) in
terms of mass-weighted averages over the system. The dots are for
$q=0.3$, 0.4, 0.6, 0.8 and 1.0, from left to right respectively. For
comparison, the dashed curve shows the relation between $h_4$ and
anisotropy for spherical constant-$\beta$ Hernquist models. The close
similarity between the results shows that it is reasonable to use
spherical models to interpret data for an ensemble of clusters that
may individually not be spherical.\label{f:axiherm}}

\newcommand{\figcapaxiRfunc}{Solid curves show, from top to bottom, 
the projected intensity, velocity dispersion and Gauss-Hermite moment
$h_4$ as function of radius for an ensemble of axisymmetric
two-integral Hernquist models of fixed axial ratio $q = 0.6$, seen
from random viewing directions. For comparison, the dashed curves in
the bottom two panels show the predictions for a spherical
constant-$\beta$ Hernquist model with the same projected intensity
profile, and the same overall anisotropy $\langle \sigma_r / \sigma_t
\rangle$. The close similarity between the results shows that it is 
reasonable to use spherical models to interpret data for an ensemble
of clusters that may individually not be spherical.\label{f:axiRfunc}}

\newcommand{\figcapgaussassump}{Gauss-Hermite moments $h_4$ and $h_6$ 
of the grand-total velocity histogram for the CNOC1 ensemble cluster,
as function of the velocity dispersion anisotropy $\sigma_r /
\sigma_t$. Solid curves show the predictions obtained with the DF
modeling approach of \S\ref{s:DFmodels} (same as the predictions in
Figure~\ref{f:gauher}, but without the shot-noise related error
bars). Dashed curves show the predictions obtained when one
incorrectly assumes that the velocity distribution is Gaussian
everywhere along the line of sight, as described in Appendix~A. The
latter assumption, which was employed in papers by Ramirez et al.,
does not yield the correct results.\label{f:gaussassump}}


\ifsubmode
\figcaption{\figcapprojnumden}
\figcaption{\figcapdispprof}
\figcaption{\figcapmassparam}
\figcaption{\figcapmassprof}
\figcaption{\figcapbackfrac}
\figcaption{\figcaplikebest}
\figcaption{\figcaphistograms}
\figcaption{\figcapgauher}
\figcaption{\figcapaxiherm}
\figcaption{\figcapaxiRfunc}
\figcaption{\figcapgaussassump}
\clearpage
\else\printfigtrue\fi

\ifprintfig


\clearpage
\begin{figure}
\centerline{\epsfbox{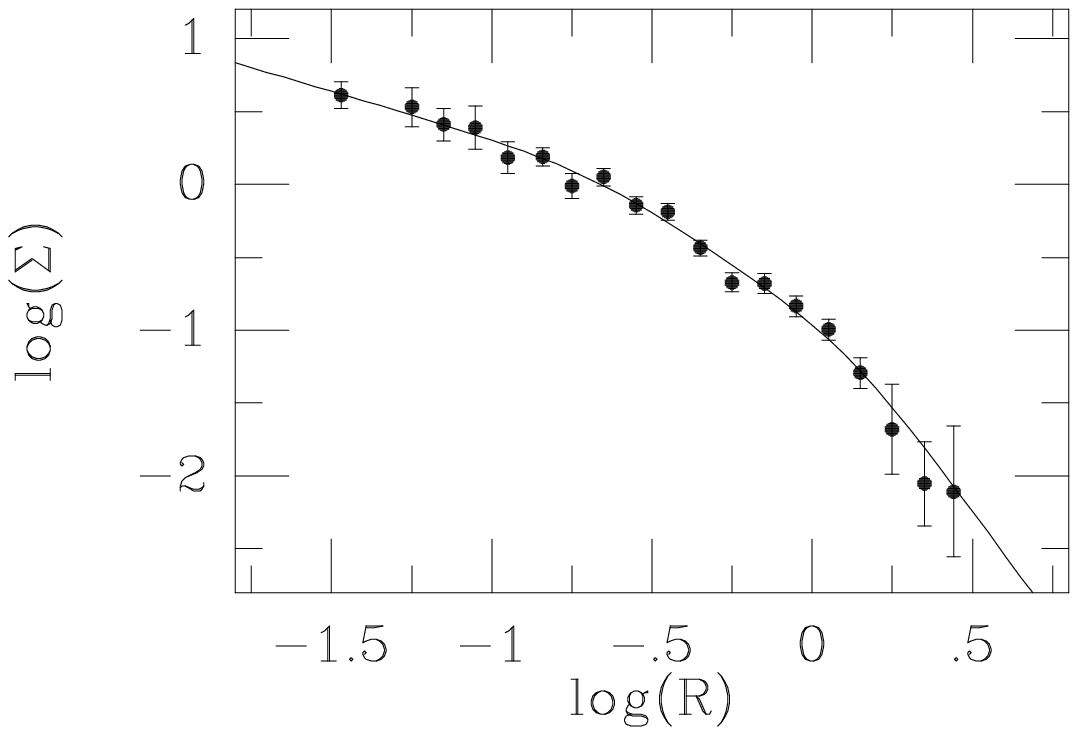}}
\ifsubmode
\vskip3.0truecm
\setcounter{figure}{0}
\addtocounter{figure}{1}
\centerline{Figure~\thefigure}
\else\figcaption{\figcapprojnumden}\fi
\end{figure}


\clearpage
\begin{figure}
\centerline{\epsfbox{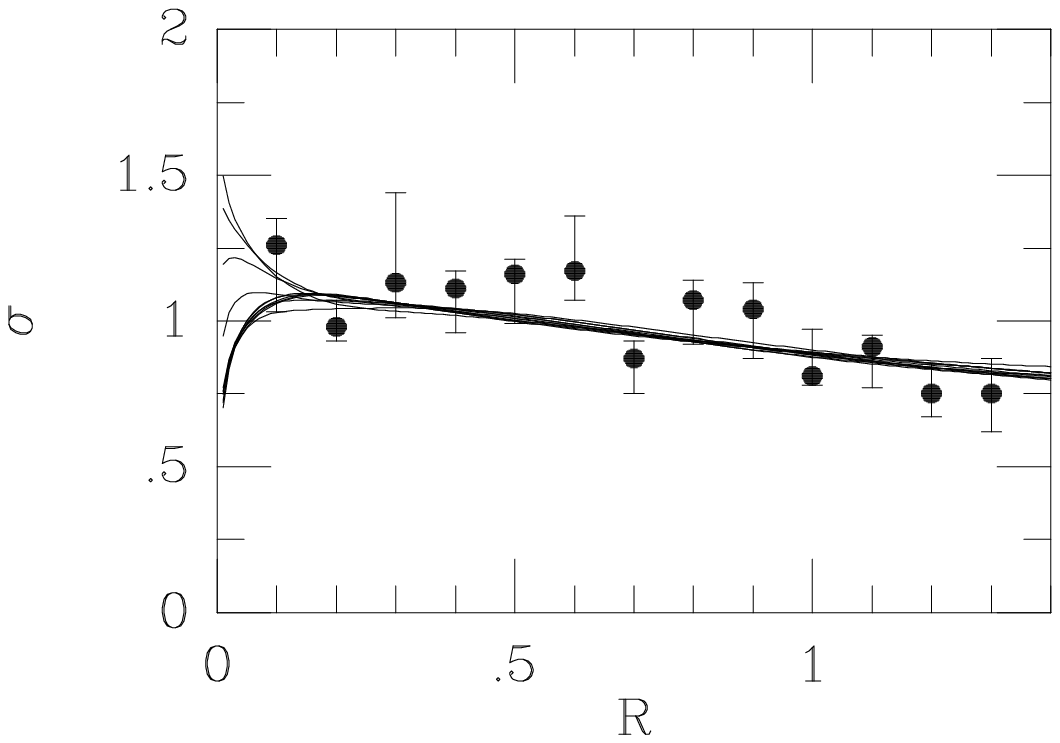}}
\ifsubmode
\vskip3.0truecm
\addtocounter{figure}{1}
\centerline{Figure~\thefigure}
\else\figcaption{\figcapdispprof}\fi
\end{figure}


\clearpage
\begin{figure}
\centerline{\epsfbox{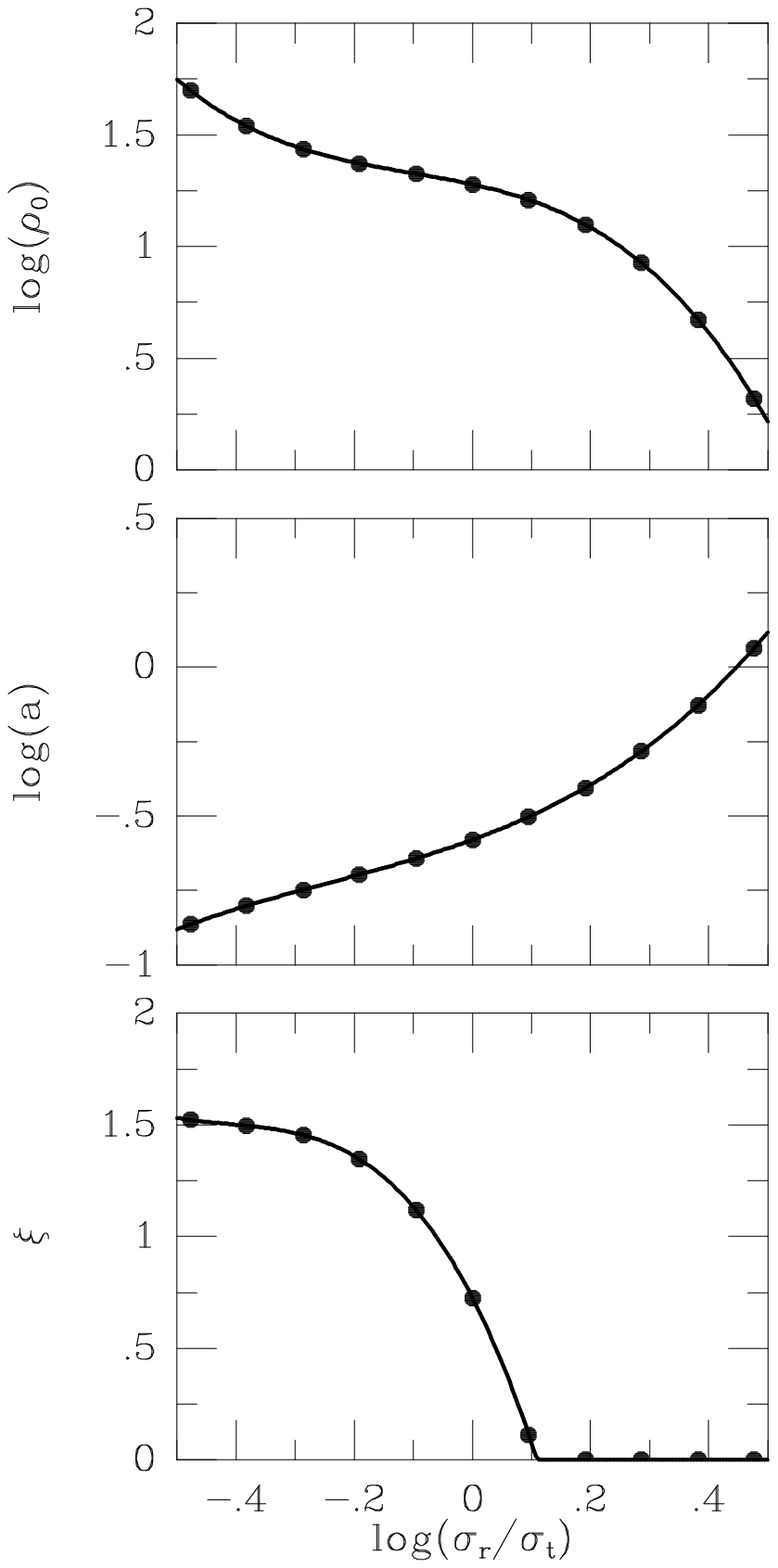}}
\ifsubmode
\vskip3.0truecm
\addtocounter{figure}{1}
\centerline{Figure~\thefigure}
\else\figcaption{\figcapmassparam}\fi
\end{figure}


\clearpage
\begin{figure}
\epsfxsize=15.0truecm
\centerline{\epsfbox{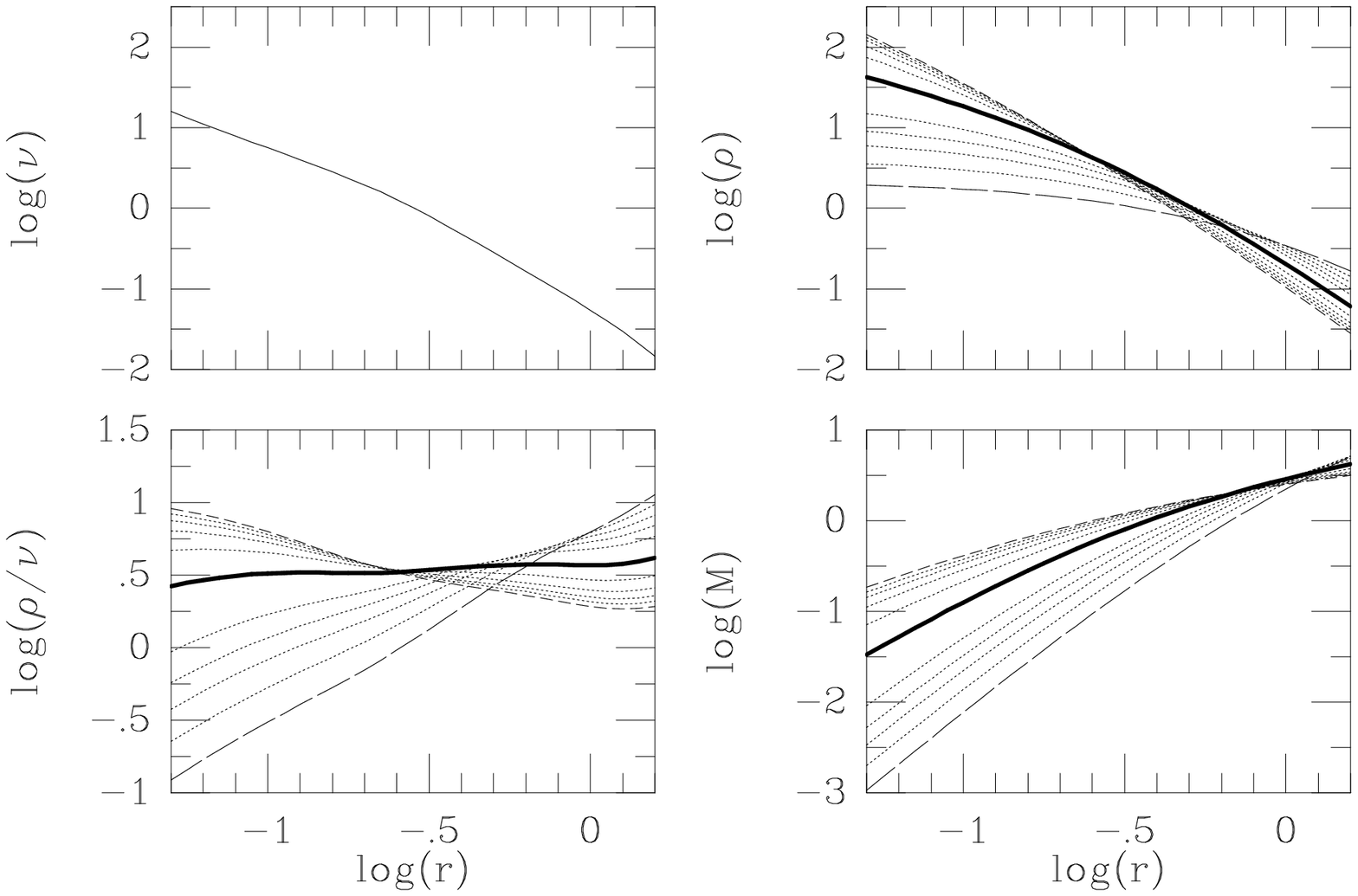}}
\ifsubmode
\vskip3.0truecm
\addtocounter{figure}{1}
\centerline{Figure~\thefigure}
\else\figcaption{\figcapmassprof}\fi
\end{figure}


\clearpage
\begin{figure}
\centerline{\epsfbox{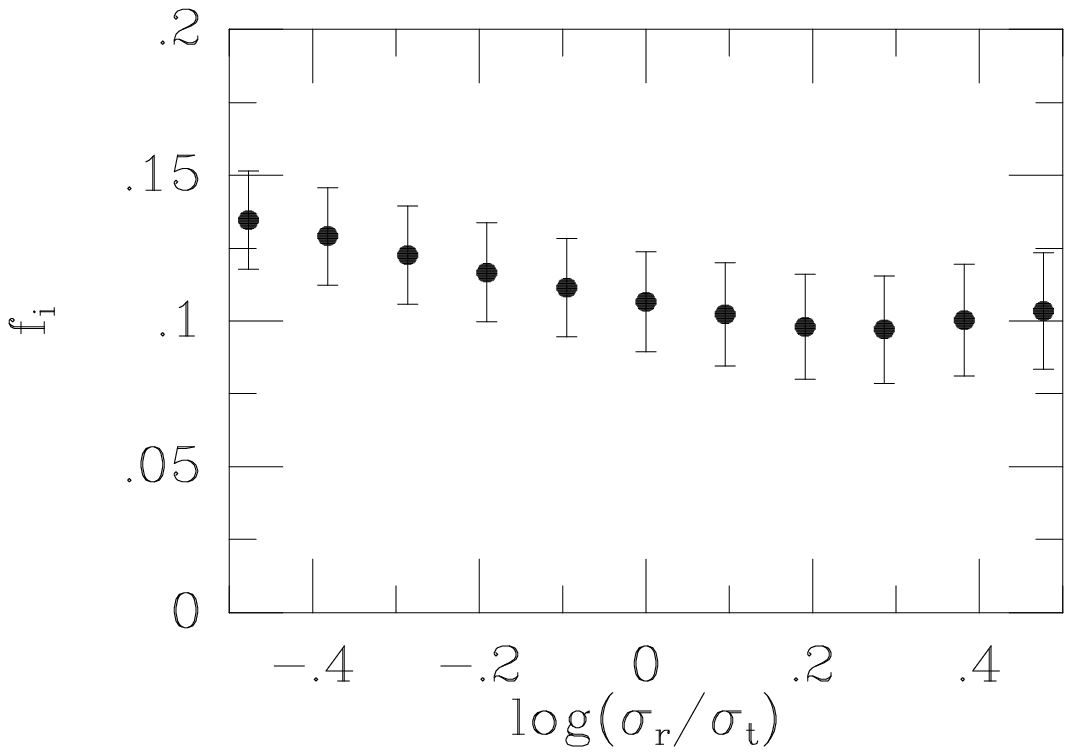}}
\ifsubmode
\vskip3.0truecm
\addtocounter{figure}{1}
\centerline{Figure~\thefigure}
\else\figcaption{\figcapbackfrac}\fi
\end{figure}
 

\clearpage
\begin{figure}
\centerline{\epsfbox{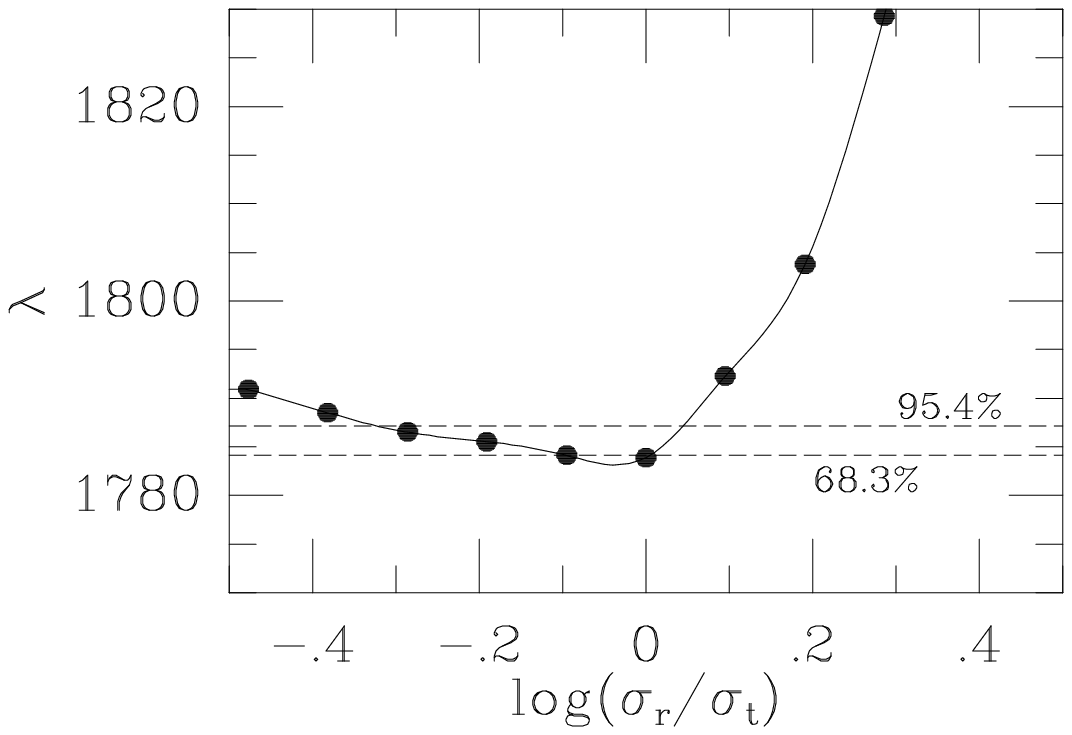}}
\ifsubmode
\vskip3.0truecm
\addtocounter{figure}{1}
\centerline{Figure~\thefigure}
\else\figcaption{\figcaplikebest}\fi
\end{figure}
 

\clearpage
\begin{figure}
\epsfxsize=15.0truecm
\centerline{\epsfbox{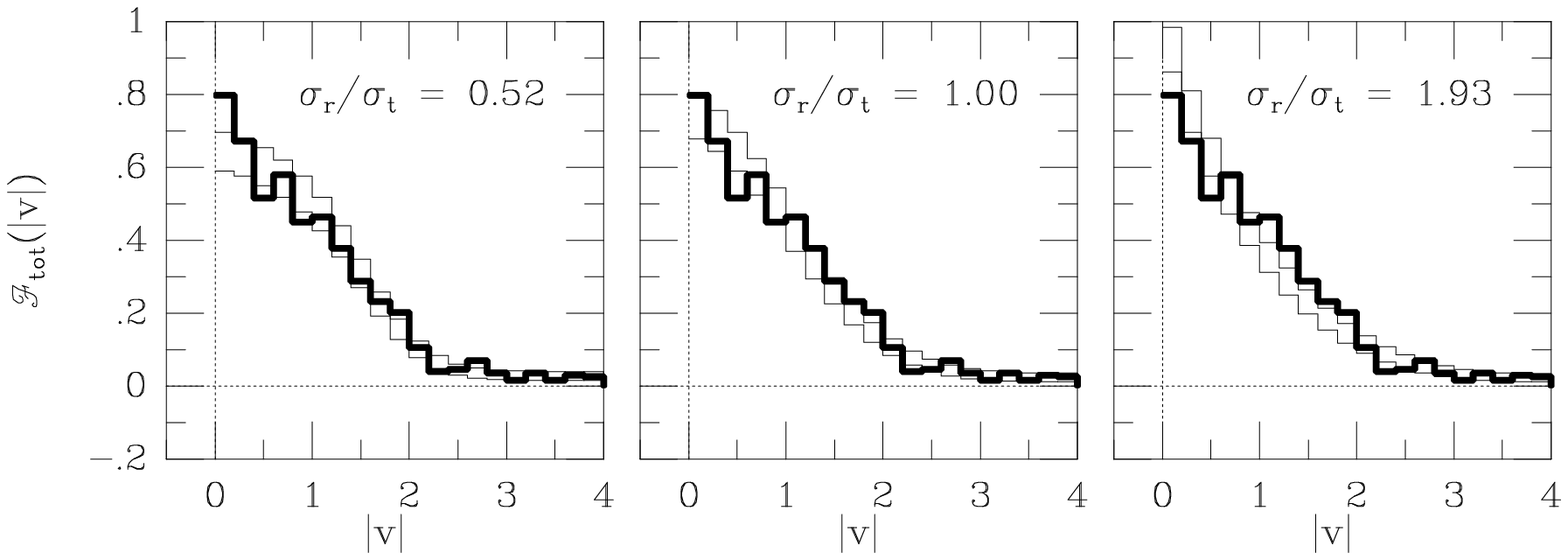}}
\ifsubmode
\vskip3.0truecm
\addtocounter{figure}{1}
\centerline{Figure~\thefigure}
\else\figcaption{\figcaphistograms}\fi
\end{figure}
 

\clearpage
\begin{figure}
\centerline{\epsfbox{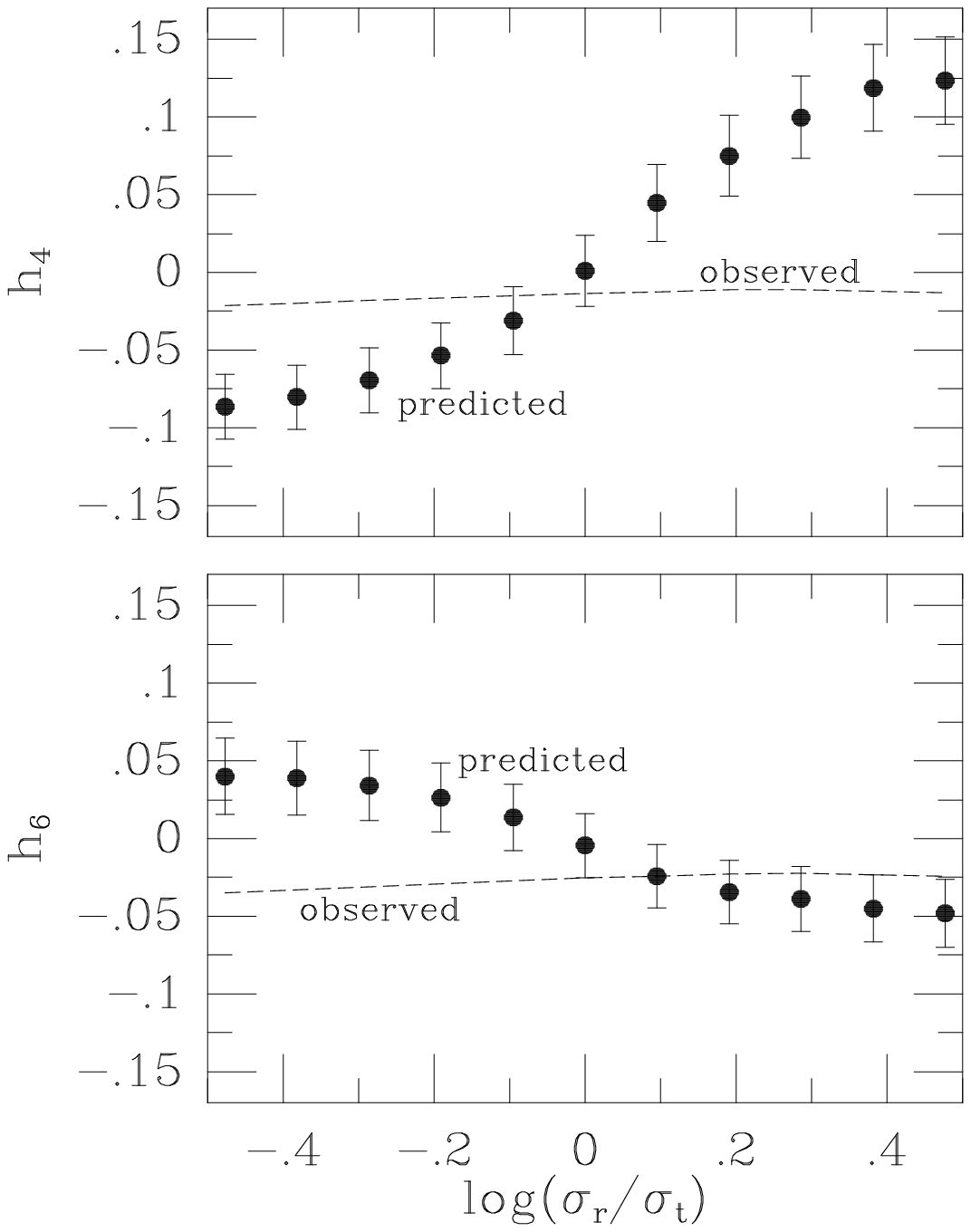}}
\ifsubmode
\vskip3.0truecm
\addtocounter{figure}{1}
\centerline{Figure~\thefigure}
\else\figcaption{\figcapgauher}\fi
\end{figure}
 
 
\clearpage
\begin{figure}
\centerline{\epsfbox{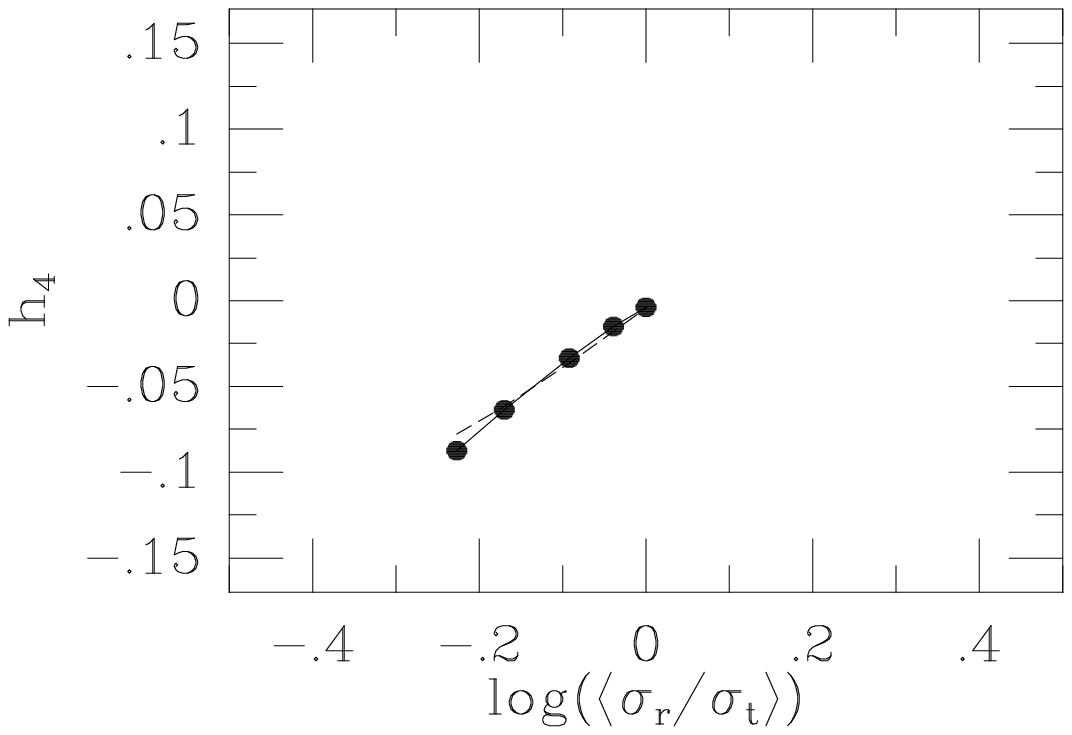}}
\ifsubmode
\vskip3.0truecm
\addtocounter{figure}{1}
\centerline{Figure~\thefigure}
\else\figcaption{\figcapaxiherm}\fi
\end{figure}
 
 
\clearpage
\begin{figure}
\epsfysize=14.0truecm
\centerline{\epsfbox{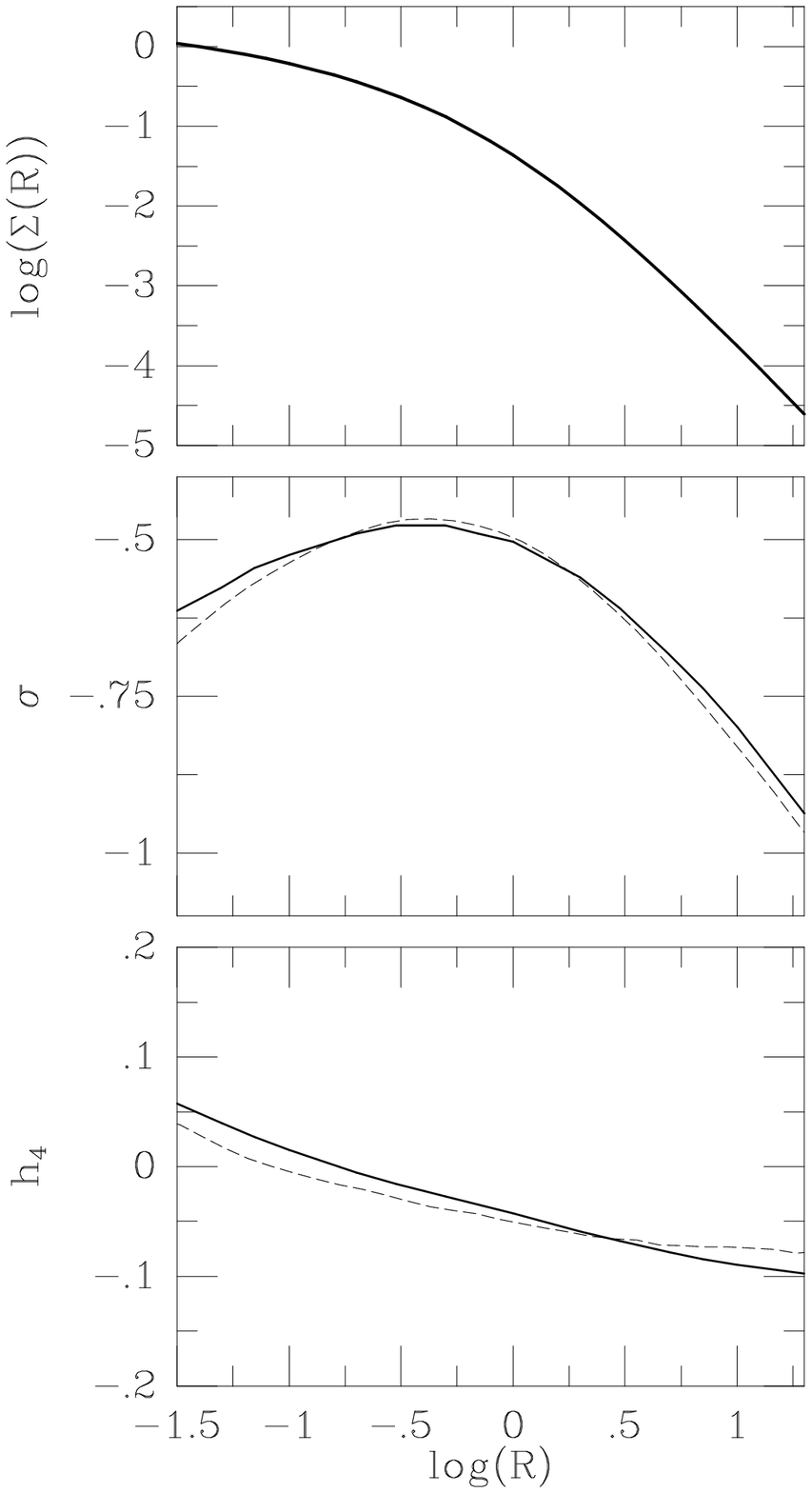}}
\ifsubmode
\vskip3.0truecm
\addtocounter{figure}{1}
\centerline{Figure~\thefigure}
\else\figcaption{\figcapaxiRfunc}\fi
\end{figure}
 
 
\clearpage
\begin{figure}
\centerline{\epsfbox{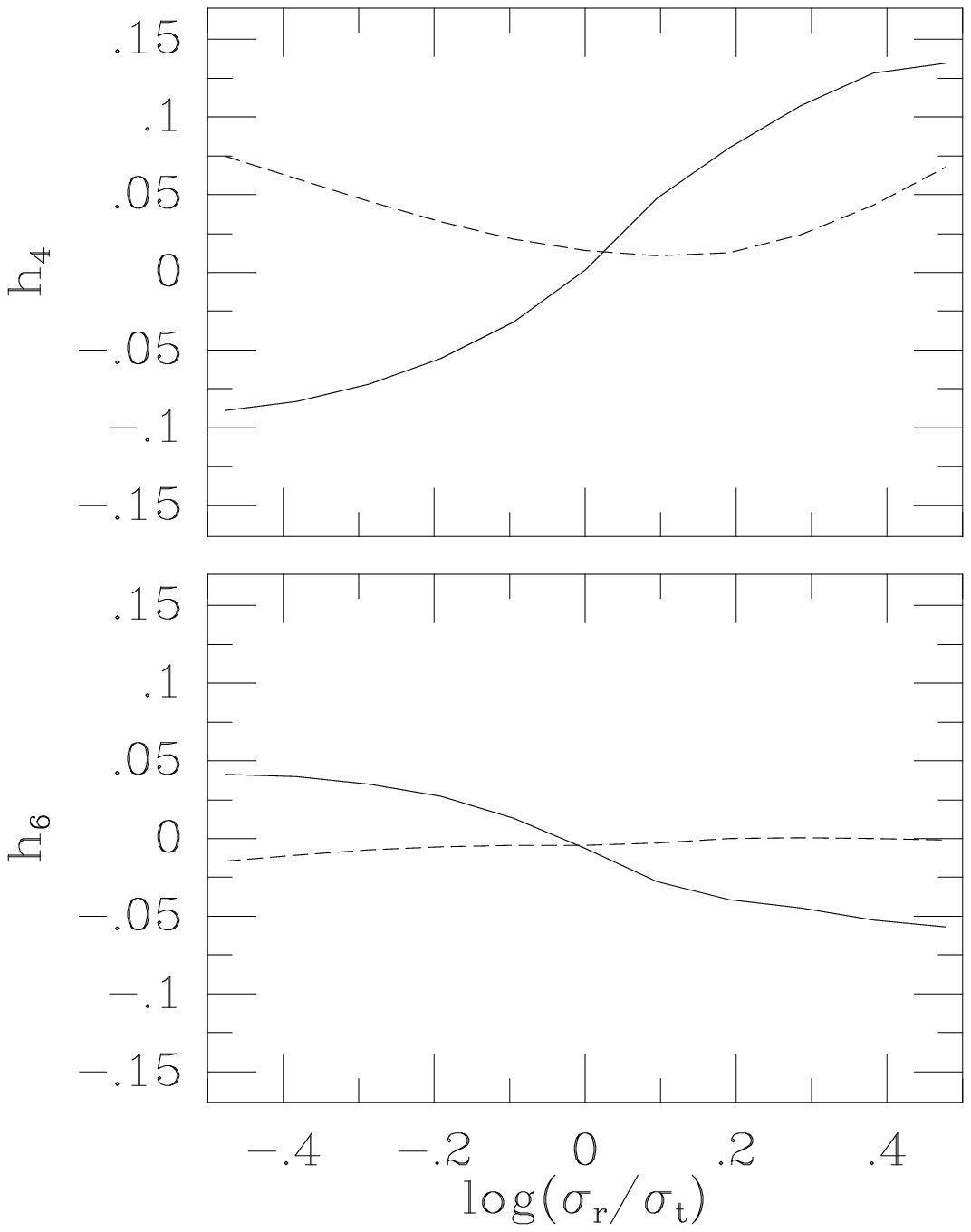}}
\ifsubmode
\vskip3.0truecm
\addtocounter{figure}{1}
\centerline{Figure~\thefigure}
\else\figcaption{\figcapgaussassump}\fi
\end{figure}
 

\fi







\end{document}